\documentclass{jfm}

\usepackage{graphicx}
\usepackage{natbib}
\usepackage{color}
\usepackage{amsfonts,amssymb,amsmath}
\usepackage{bm}
\usepackage{multirow}

\graphicspath{{../../figures/}}

\ifCUPmtlplainloaded \else
  \checkfont{eurm10}
  \iffontfound
    \IfFileExists{upmath.sty}
      {\typeout{^^JFound AMS Euler Roman fonts on the system,
                   using the 'upmath' package.^^J}%
       \usepackage{upmath}}
      {\typeout{^^JFound AMS Euler Roman fonts on the system, but you
                   dont seem to have the}%
       \typeout{'upmath' package installed. JFM.cls can take advantage
                 of these fonts,^^Jif you use 'upmath' package.^^J}%
      }
  \else
  \fi
\fi

\ifCUPmtlplainloaded \else
  \checkfont{msam10}
  \iffontfound
    \IfFileExists{amssymb.sty}
      {\typeout{^^JFound AMS Symbol fonts on the system, using the
                'amssymb' package.^^J}%
       \usepackage{amssymb}%

      }{}
  \fi
\fi

\ifCUPmtlplainloaded \else
  \IfFileExists{amsbsy.sty}
    {\typeout{^^JFound the 'amsbsy' package on the system, using it.^^J}%
     \usepackage{amsbsy}}
    {\providecommand\boldsymbol[1]{\mbox{\boldmath $##1$}}}
\fi

\newcommand\Rey{\mbox{\textit{Re}}}  

\newsavebox{\astrutbox}
\sbox{\astrutbox}{\rule[-5pt]{0pt}{20pt}}

\newcommand{\BS}[1]{{\color{black}#1}}
\newcommand{\xu}[1]{{\color{black}#1}}

\title[Turbulent fronts in pipe flow]{Propagation speed of turbulent fronts in pipe flow at high Reynolds numbers}

\author[Kaiwen Chen, Duo Xu, Baofang Song]%
{Kaiwen Chen$^1$,
Duo Xu$^{2, 3}$,
Baofang Song$^1$
\thanks{Email address for correspondence: baofang\_song@tju.edu.cn}}
\affiliation{$^1$ Center for Applied Mathematics, Tianjin University, Tianjin 300072, China\\[\affilskip]
$^2$ The State Key Laboratory of Nonlinear Mechanics, Institute of Mechanics, Chinese Academy of Sciences, Beijing 100190, China\\[\affilskip]
$^3$ School of Engineering Science, University of Chinese Academy of Sciences, Beijing 100049, China\\[\affilskip]
}

\date{?; revised ?; accepted ?. - To be entered by editorial office}
\begin{document}

\maketitle

\begin{abstract} 
We investigated the \xu{propagation} of turbulent fronts in pipe flow at high Reynolds numbers by direct numerical simulation. 
We used a technique combining a moving frame of reference and an artificial damping to isolate the fronts in short periodic pipes, which enables us to explore the bulk Reynolds number up to $Re=10^5$ with affordable computation power. We measured the propagation speed of the downstream front and observed that a fit of $1.971-(Re/1925)^{-0.825}$ (in unit of bulk speed) well captures this speed above $Re\simeq 5000$. The speed increases monotonically as $Re$ increases, in stark contrast to the decreasing trend above $Re\simeq 10000$ reported by \citet{Wygnanski1973}. The speed of the upstream front overall agrees with the former studies and $0.024+(Re/1936)^{-0.528}$ well fits our data and those from the literature. Based on our analysis of the front dynamics, we proposed that both front speeds would keep their respective monotonic trends as the Reynolds number increases further. We show that, at high Reynolds numbers, the local transition at the upstream front tip is via high-azimuthal-wavenumber structures in the high-shear region near the pipe wall, whereas at the downstream front tip is via low-azimuthal-wavenumber structures in the low-shear region near the pipe center. This difference is possibly responsible for the asymmetric speed scalings between the upstream and downstream fronts.
\end{abstract}

\begin{keywords}
\end{keywords}

\section{Introduction}
At sufficiently high Reynolds numbers, once perturbed locally, pipe flow becomes turbulent and develops two turbulent fronts on upstream and downstream \citep{Lindgren1957,Wygnanski1973}. The turbulent region expands via the propagation of the two fronts, and therefore, the dynamics of the fronts determines how the laminar flow is entrained into the turbulent region and how fast the turbulent region can expand. Since \citet{Lindgren1957}, many studies are devoted to studying turbulent fronts in straight and bent pipes in the past decades \citep{Wygnanski1973,Lindgren1969,Durst2006,Nishi2008,Duguet2010b,Holzner2013, Barkley2015,Song2017,Schlatter2019}, among which the global propagation speed of the fronts has been an important subject. 

A few studies attempted to theoretically derive the front speed. For the upstream front (UF), by analysing the energy flux across a control volume enclosing the entire front region (the part between the parabolic laminar flow on the upstream and fully developed turbulent flow on the downstream), \citet{Lindgren1969} derived an asymptotic speed of 0.69 in theory, and speculated 0.64 being a lower limit of the speed in experiments as $\Rey\to\infty$. However, as \citet{Wygnanski1973} pointed out, these values are rather far away from experimental measurements at high Reynolds numbers, and a trend approaching these values was not supported by the measurements. Treating the front as an isosurface of a proper quantity (e.g. the \xu{enstrophy}), \citet{Holzner2013} theoretically derived the local speed of the three dimensional isosurface relative to the local flow speed and evaluated the contributions from different physical mechanisms. Given the local flow speed, one can in principle calculate an instantaneous axial propagating speed of the isosurface as a whole. However, as the authors pointed out, the method requires fully resolving the complex highly-convoluted isosurfaces and is challenging to implement for both numerical simulations and experiments. Besides, the method requires detailed local flow speed at the isosurface to calculate the global propagation speed of the isosurface.

Regarding the instability and transition mechanisms at the fronts, several qualitative mechanisms were proposed for the upstream front of turbulent puffs ({\color{black} localised turbulence with an approximately constant axial extension} ) and slugs ({\color{black} turbulent structures expanding towards both upstream and downstream directions along the pipe axis}) at low and moderate Reynolds numbers. For example, \citet{Shimizu2009} and \citet{Duguet2010b} proposed the Kelvin-Hemholtz instability, while \citet{Hof2010} proposed an inflectional instability associated with the low-speed streaks. However, to our knowledge, there still lacks consensus and quantitative understanding of the mechanism. As a result, predicting the front speed theoretically from a first principle has not been realised, and determining the propagation speed by tracking the front remains the main approach in either experiments or numerical simulations.

\begin{figure}
\centering	
 \includegraphics[width=0.8\linewidth]{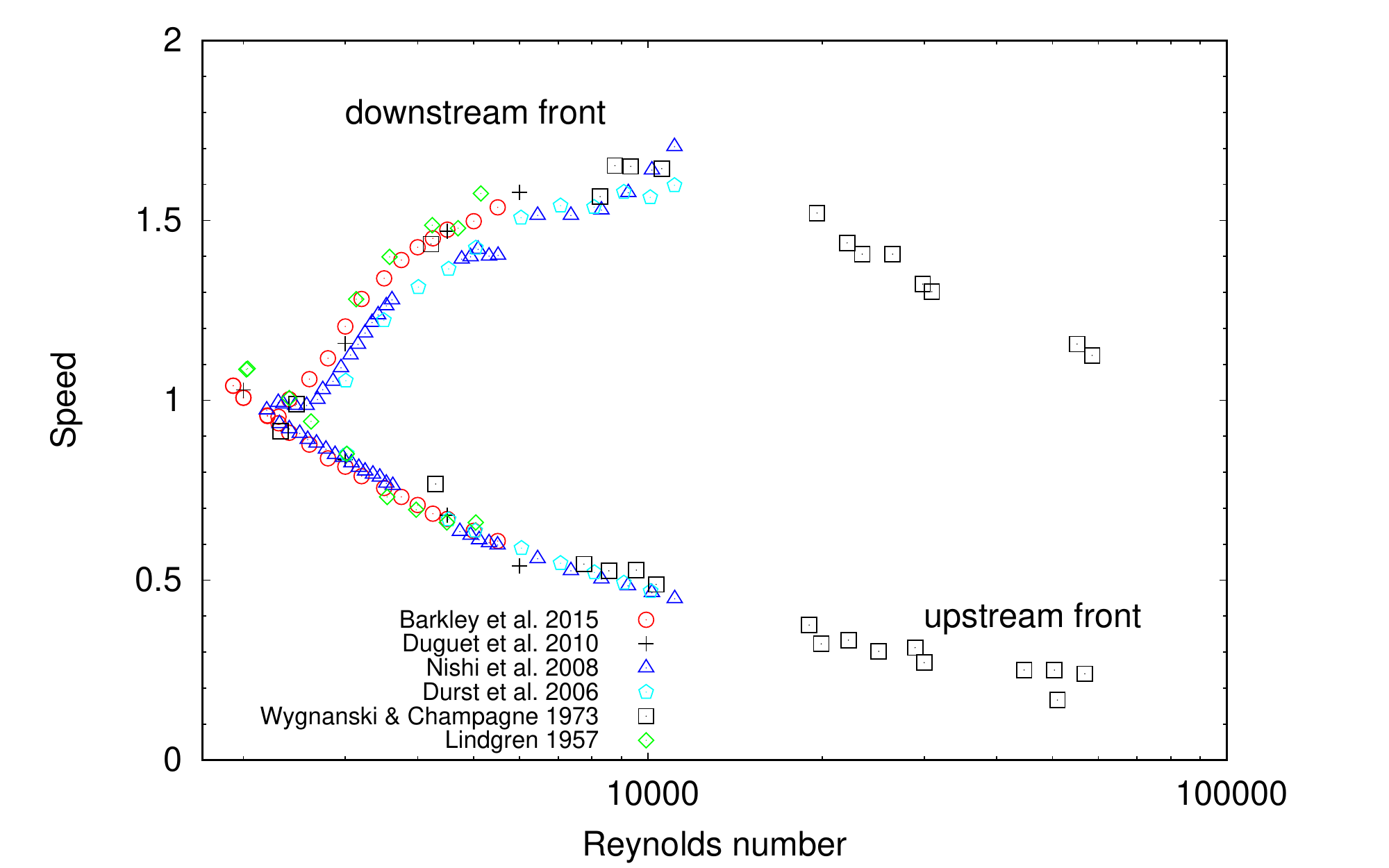}
\caption{\label{fig:front_speed_literature} Global axial propagation speed as a function of the bulk Reynolds number taken from the literature.}
\end{figure}

Figure \ref{fig:front_speed_literature} presents some important studies on the global propagation speed of the fronts, in which the bulk Reynolds number is defined as $\Rey=UD/\nu$ with $U$ being the bulk speed (axial velocity averaged in the pipe cross-section), $D$ the pipe diameter and $\nu$ the kinematic viscosity of the fluid. We use this definition of the Reynolds number, and normalise length by $D$ and speed by $U$ in this paper. It can be seen that the data sets for the UF speed roughly agree with each other, whereas show much scattering for the downstream front (DF) speed. What is more, relevant study at high Reynolds numbers ($\gtrsim$ 10000) is rare. 

Among the studies, \citet{Wygnanski1973} carried out a comprehensive investigation into the kinematics and structure of the fronts, and considered the highest Reynolds number so far (up to $\Rey\simeq 60000$). In fact, to our knowledge, their work was the only one that covers the regime of $\Rey \gtrsim 10000$. Based on their measurements, the authors concluded that the speeds of both the UF and DF decrease as Reynolds number increases above $\Rey\simeq 10000$ such that the difference between the two, i.e. the expansion rate of the turbulent region, approaches the bulk speed $U$ as $\Rey\to\infty$ (see figure \ref{fig:front_speed_literature}). However, {\color{black} the basic laminar flow had not fully developed in their pipe at high Reynolds numbers (indicated by their figure 9). Therefore, the front speeds they reported at high Reynolds numbers were likely affected by the insufficiently developed basic flow, although }they didn't explain the mechanism responsible for the decreasing trend of the DF speed above $\Rey\simeq 10000$.
\xu{\citet{Barkley2015} and \citet{Barkley2016}} presented a state-of-the-art study on the fronts in the transitional Reynolds number regime ($\Rey<6000$) by combining experiments, numerics and theoretical modelling. Their advection-reaction-diffusion model well describes the front dynamics and kinematics on large-scale in the transitional regime. With an assumption that \BS{the structure of} the UF and DF gradually \xu{becomes} the mirror image of each other as $\Rey$ increases, \BS{based on one-dimensional representations of the flow along pipe axis \citep{Nishi2008,Duguet2010b,Song2017}}, they predicted that the front speeds will be antisymmetric about the advection speed of the bulk turbulence (referred to as the neutral speed in \citet{Barkley2015}). Their prediction implied that the DF speed would monotonically increase as $\Rey$ increases,  conflicting with the measurements of \citet{Wygnanski1973}. However, their model analysis was only informed by data (both experimental and numerical) up to $\Rey\simeq 6000$, therefore, it is unclear if the model quantitatively describes the fronts at high Reynolds numbers.

In a word, there has been no consensus on the kinematics of the DF at high Reynolds numbers. One difficulty in measuring the front speed at high Reynolds numbers in experiments, especially for the DF, is that the speed takes a long time to saturate given usual initial perturbations (e.g. transverse jets and impulsive partial blockage) so that the pipe should be sufficiently long for the DF to develop. \citet{Nishi2008} used a 533$D$ (8~m) pipe and found that the DF speed experiences long transients and the higher the Reynolds number, the longer the transient. They only performed measurements up to $\Rey=11000$. \citet{Wygnanski1973} used a 500$D$ pipe to measure the front speed up to $Re\simeq 60000$. However, they didn't show the development of the front speed with time, and therefore, whether or not the front speed had saturated is not clear {\color{black} (it probably had not at high Reynolds numbers because the basic flow was still developing at their measurement point)}.
A second difficulty, which is possibly more difficult to overcome, is that small \xu{environmental} disturbances may trigger turbulence at high Reynolds numbers and affect the fronts \citep{Hof2003,Peixinho2007}.  {\color{black} Indeed, \citet{Wygnanski1973} noticed increasing low-frequency oscillations as $\Rey$ increases in their basic flow, which may also contribute to the deviation of the basic flow from the parabolic laminar flow.}

In numerical simulations, \xu{environmental} disturbances can be kept low and periodic pipes can keep the front from exiting the pipe, however, the fast growth of slugs at high Reynolds numbers quickly renders the pipe fully turbulent if the pipe is not sufficiently long. \citet{Song2014} (see their \xu{figure} 3-5) reported that, by using localised puffs as the initial perturbation (the turbulent core is about 10$D$-long),  the pipe length needed to obtain a saturated DF speed increases as $\Rey$ increases. For example, at $\Rey=3000$, the DF speed saturates as slugs grow to a length of about 30$D$, whereas this length grows nearly linearly to $80D$ at $\Rey=4250$ and to $125D$ at $\Rey=5500$. Therefore, the pipe length should be significantly longer than 125$D$ for $\Rey=5500$ (\citet{Song2014} used a 180-$D$ pipe). At higher Reynolds numbers, the pipe length, consequently the computational cost, grows rapidly. As an estimate, for DNS that uses uniform grids in the axial and azimuthal directions, a 250$D$-long pipe for $Re=10000$ needs approximately half a billion grid points, and this number grows to approximately ten billion for $Re=40000$. Besides the increasing number of grid points,
the decreasing time-step size and increasing time for the speed to saturate as $\Rey$ increases also make the DNS study in the normal approach infeasible.

In this work, we measure the global axial \xu{propagation} speed of the fronts using a technique combining a moving frame of reference and an artificial damping in relatively short periodic pipes to isolate individual fronts. Besides, by using the well-developed fronts simulated at close Reynolds numbers as initial conditions, the initial adjustment of the flow can be drastically shortened. These strategies circumvent the aforementioned difficulties and enable us to investigate turbulent fronts at unprecedentedly high Reynolds numbers using DNS. 

\section{Methods}\label{sec:methods}
\subsection{Numerical methods}
We solve the incompressible Navier-Stokes equations in a moving frame of reference with a speed of $c$ and an artificial damping term with the form of $-\beta(z)\boldsymbol u$, i.e. 
\begin{equation}\label{N-S}
 \frac{\partial\boldsymbol u}{\partial t}+{(\boldsymbol u+c)}\cdot\boldsymbol{\nabla}
{\boldsymbol u}=-{\boldsymbol{\nabla}p}+\frac{1}{Re}\nabla^2{\boldsymbol u} - \beta(z)\boldsymbol u, \;
\hspace{3mm}\boldsymbol{\nabla}\cdot{\boldsymbol u}=0,
\end{equation}
where $\boldsymbol u$ is the velocity with respect to the moving frame of reference and $p$ the pressure. {\color{black}The volume flux (the bulk speed) is kept constant during the simulation}. The form of the damping term is inspired by \citet{Kanazawa2018}. The equations are solved in cylindrical coordinates $(r,\theta,z)$, which represent radial, azimuthal and streamwise coordinates, respectively. We use the open-source pipe flow code OPENPIPEFLOW \citep{Willis2017, Willis2009} to perform the simulations. As $\beta$ is dependent on $z$, the damping term is treated as a \xu{nonlinear} term in the time-stepping. The resolutions for all simulations performed in this paper \xu{are} listed in Appendix \ref{sec:appendix}.

\begin{figure}
\centering
\includegraphics[width=0.95\linewidth]{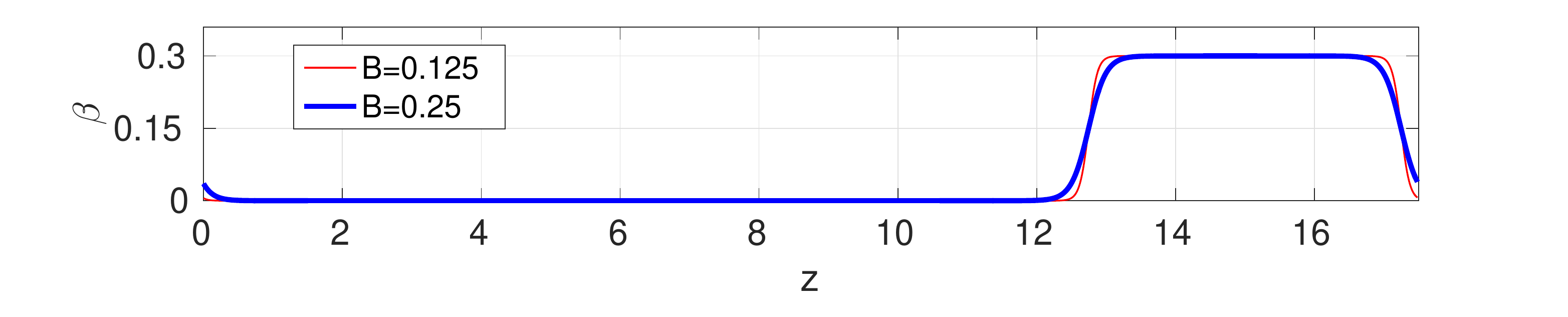}
\caption{\label{fig:damping_coefficient} The damping coefficient $\beta$ with $A$=0.3, $z_0=15$ and $R=2.25$. The thin red line is for $B=0.125$ and the bold blue line is for $B=0.25$. Note that periodic boundary condition is considered for $\beta$.}
\end{figure}

The moving frame of reference and damping term play the role of tracking and isolating a front or even a part of a front in a short pipe domain, such that we can explore the high $\Rey$ regime. This strategy is only justified if the front is locally self-sustained and does not depend on the flow far from it. This is true for the fronts of strong slugs according to \citet{Barkley2015}, \citet{Barkley2016} and \citet{Song2017}, and will also be evidenced later in this paper. {\color{black} The DF of puffs and weak slugs at low Reynolds numbers is not self-sustained, to which our technique may not apply.} The damping coefficient $\beta$ is a function of only the axial coordinate and will be used to confine the damping effect in only a part of the pipe. Specifically, we choose the following form,
\begin{equation}
\label{equ:damping_coefficient}
\beta(z)=A\left(0.5-0.5\tanh{\left(\frac{|z-z_0|-R}{B}\right)}\right),
\end{equation}
where $A$ is the amplitude, $R$ the nominal half-width of the damping region and $B$ controls the steepness of the decay of the damping at the boundary of the damping region. Therefore, this coefficient localises the damping roughly in a region of $z_0-R<z<z_0+R$. Ideally, the speed of the moving frame $c$ should be set to the speed of the front, which however is not known {\it a priori}. Therefore, $c$ is first estimated (e.g. using the front speed measured at a close $\Rey$) and then adjusted \xu{dynamically} in run time, so that the axial location of the front does not change significantly. Figure \ref{fig:damping_coefficient} shows the shape of $\beta$ given $A=0.3$, $z_0=15$ and $R=2.25$ with $B=0.25$ (the thin red line) and with $B=0.125$ (the bold blue line).

\section{Speed measurements}\label{sec:speed_measurements}
\subsection{Validation}
First, we validate our technique by comparing the front speed for $\Rey=5000$ with the data of \citet{Barkley2015} and \citet{Song2017} which were obtained in long stationary periodic pipes. The system setting is characterised by the pipe length $L$, a reference position of the front $z_{f0}$, and damping parameters $z_0$, $R$, $A$ and $B$ as defined in (\ref{equ:damping_coefficient}). For locating the front in the axial direction, following \citet{Song2017}, we set a threshold in the cross-sectional kinetic energy,
\begin{equation}
\label{equ:q}
q(z):=\int_0^1\int_0^{2\pi}(u_r^2+u_\theta^2)r d\theta dr,
\end{equation}
above which the flow is considered sufficiently turbulent. {\color{black} As \cite{Song2014} and \cite{Song2017} reported, the specific value of this threshold would not affect the average front speed as long as it is in a reasonable range.}

\begin{figure}
\centering
\includegraphics[width=0.95\linewidth]{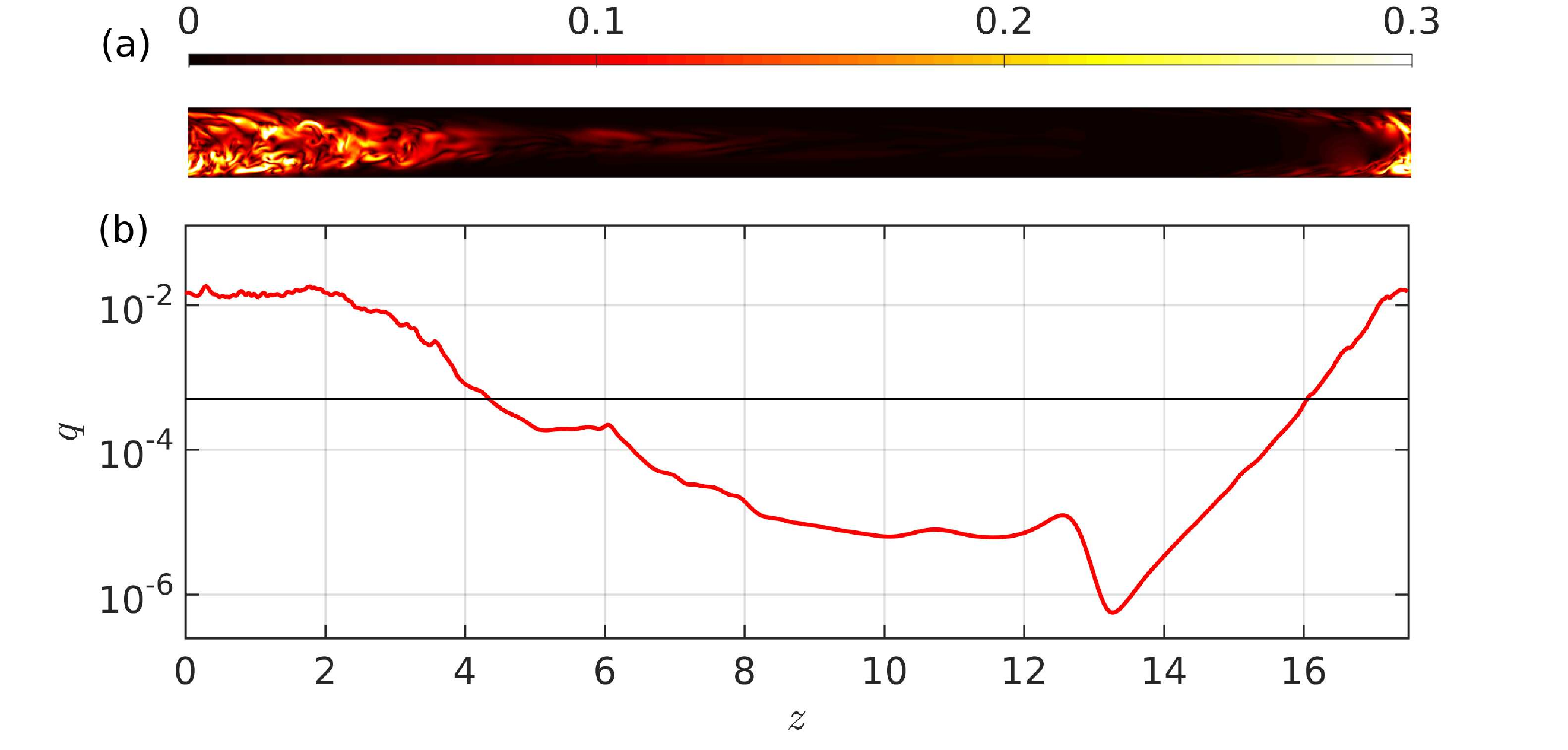}
\caption{\label{fig:crosssec0180_Re5000} The parameter setting of $L=17.5$, $z_{f0}=5$, $z_0=15$, $R=2.25$, $A=0.3$ and $B=0.25$ for $\Rey=5000$. The estimated front speed $c_0=1.50$. The damping coefficient $\beta$ with these parameters are exactly the one shown in figure \ref{fig:damping_coefficient}. (a) One snapshot of the front in the $r$-$z$ cross-section, in which the transverse velocity $\sqrt{u_r^2+u_\theta^2}$ is colour-coded. The front is on the left (upstream) and the damping region is on the right (downstream), with the laminar gap in between (the dark region). (b) The quantity $q$ defined as (\ref{equ:q}) plotted along the pipe axis at a time instant (red bold line). The horizontal black thin line marks the threshold by which the location of the front is determined. The nominal axial position of the front is $z_f=4.3$ at this time instant, given by the left intersection between the red curve and black line.}
\end{figure}

\begin{figure}
\centering
\includegraphics[width=0.99\linewidth]{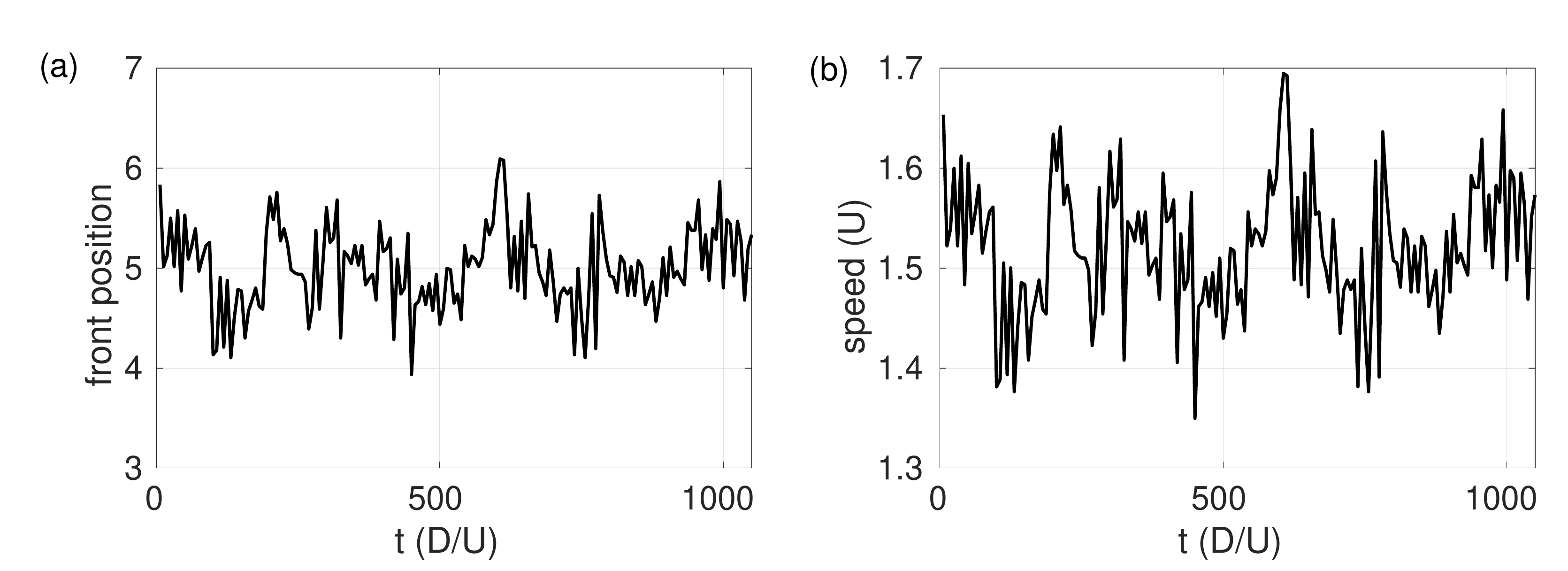}
\caption{\label{fig:front_location_front_speed} (a) The nominal axial position of the front, determined by the threshold in $q$ as described in the caption of figure \ref{fig:crosssec0180_Re5000}, for the DF at $Re=5000$. (b) The speed of the frame of reference as a function of time.}
\end{figure}

Figure \ref{fig:crosssec0180_Re5000} shows a set-up for the DF at $\Rey=5000$ with a pipe length $L=17.5$ (see the parameters in table \ref{tab:parameter_setting}). We aim to keep the axial location of the front at around $z_{f0}=5$. The damping coefficient $\beta$ is set as that shown in figure \ref{fig:damping_coefficient}. This set-up gives approximately a $7D$-long laminar gap between the front and the damping region. Figure \ref{fig:crosssec0180_Re5000}(a) shows a snapshot of the front in the $r$-$z$ cross-section, and figure \ref{fig:crosssec0180_Re5000}(b) shows $q(z)$ along the pipe axis. With an estimation of the front speed $c_0=1.50$, the axial location of the front $z_f$ is determined using the threshold $5\times 10^{-4}$ in $q$ approximately every $\delta t=6.25D/U$, see figure \ref{fig:front_location_front_speed}(a).  Subsequently the speed of the frame of reference $c$ is \xu{dynamically} adjusted according to the location of the front as
\begin{equation}\label{equ:speed_adjustment}
c+(z_f-z_{0f})/\delta t\to c,
\end{equation}
so that the front does not move too far away from the reference position $z_{f0}=5$, see figure \ref{fig:front_location_front_speed}. {\color{black} Note that other values of $\delta t$ can be chosen as long as the position of the front does not fluctuate much.} As shown in the figure, this technique indeed can isolate the front and track it for a very long time (a time window of approximately 1050 $D/U$ is shown in the figure) in the $17.5D$ pipe. 

\begin{figure}
\centering
\includegraphics[width=0.99\linewidth]{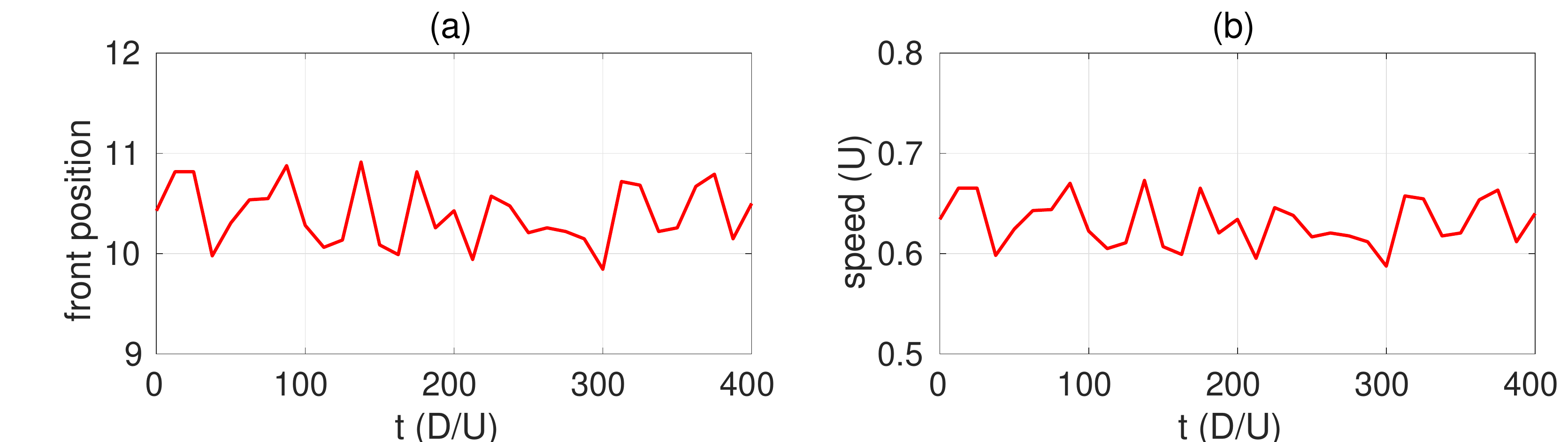}
\caption{\label{fig:front_location_front_speed_UF} (a) The nominal axial position of the UF with the threshold of $5\times 10^4$ in $q$ (see table \ref{tab:parameter_setting}) at $Re=5000$ in the $17.5D$ pipe. (b) The speed of the frame of reference.}
\end{figure}

As the location of the front only fluctuates slightly around $z_{f0}=5$ (within 1$D$) in the moving frame of reference over a long time, we can approximate the front speed as the average speed of the frame of reference. In this test, the average of $c$ is about 1.518. To show the effect of the pipe length to the front speed, we also measured the speed in a $35D$ pipe with the parameters shown in table \ref{tab:parameter_setting}, which gives a laminar gap of approximately 14$D$.  
The measurement gives 1.520 averaged over $1250 D/U$, which is very close to the speed measured in the $17.5D$ pipe. Therefore, it seems that the speed measured in our simulations is not significantly affected by the pipe length as well as the spatial extension and strength of the damping. The speed of the DF measured in a 180$D$ pipe by \citet{Barkley2015} and \citet{Song2017} is 1.498, which agrees with our current result within an error of about 1.3\%. The speed of the UF for $\Rey=5000$ is also measured in the $L=17.5D$ pipe with the parameters shown in table \ref{tab:parameter_setting}. The speed is 0.630 averaged over $400D/U$ (see figure \ref{fig:front_location_front_speed_UF}), while \citet{Barkley2015} and \citet{Song2017} reported 0.637 in a $180D$ pipe. The speed was measured in a shorter time window than the DF because \citet{Nishi2008} and \cite{Song2014} both reported that the speed of the UF stabilises much more quickly and fluctuates much more weakly than that of the DF, which was indeed also observed in the present work (to compare the fluctuations of the front position and frame speed in figure \ref{fig:front_location_front_speed} and figure \ref{fig:front_location_front_speed_UF}). Table \ref{tab:parameter_setting} summarises the parameters and measured front speeds of these tests. 

These tests suggest that this technique indeed enables us to study turbulent fronts in short periodic pipes without significant domain size effect. Besides serving as a validation, the results shown here also suggest that the front is indeed locally self-sustained and does not depend on the turbulent flow far from it, as the front speed does not change if the turbulence sufficiently far from it is damped. It should be noted that, the total observation time ($1050D/U$) of the well-developed DF in this $17.5D$ pipe based on a single run is roughly the same as the sum of the observation time of \citet{Song2014} based on 20 runs with different puffs as initial conditions in a $180D$ pipe. In comparison, the computational cost is greatly reduced by using this technique.

\begin{table}
\renewcommand\arraystretch{1.2}
\centering
\begin{tabular}{p{0.1cm}p{0.4cm}p{0.7cm}p{0.7cm}p{0.7cm}p{0.5cm}p{0.7cm}p{0.7cm}p{0.7cm}p{1.5cm}p{2.0cm}p{0.7cm}p{0.7cm}}
& & ${\Rey}$ & ${L}$ & $z_{f0}$	& $z_0$	&  $R $	& $A$	& $B$	& threshold  & \text{max}$\Delta t$  & $T$ & {speed} \\
\hline
\multirow{2}{*}{\rotatebox[origin=c]{90}{DF}} & \multirow{2}{*}{\rotatebox[origin=c]{90}{}}  
 & 5000	& 17.5	& 5	  & 15	& 2.25	& 0.3	& 0.25	& $5\times 10^{-4}$ & $2.0\times 10^{-3}$   & 1050	& 1.518\\
& & 5000	& 35	& 11.5	& 29	& 3.5	& 0.1	& 0.5	& $5\times 10^{-4}$ & $2.0\times 10^{-3}$ & 1250	& 1.520\\
\hline

\multirow{1}{*}{\rotatebox[origin=c]{90}{UF}} & \multirow{1}{*}{\rotatebox[origin=c]{90}{}} 
& 5000 	& 17.5	& 10.5	& 5.0	& 4.5	& 0.1	& 0.25	& $5\times 10^{-4}$ & $1.25\times 10^{-3}$ & 400	& 0.630 \\
\end{tabular}

\caption{\label{tab:parameter_setting} Pipe length, damping parameter setting, threshold in $q$, time-step size and averaging time of the speed for the DF at $\Rey=5000$.}
\end{table}

\subsection{Fronts at high Reynolds numbers}
\begin{figure}
\centering
\includegraphics[width=0.99\linewidth]{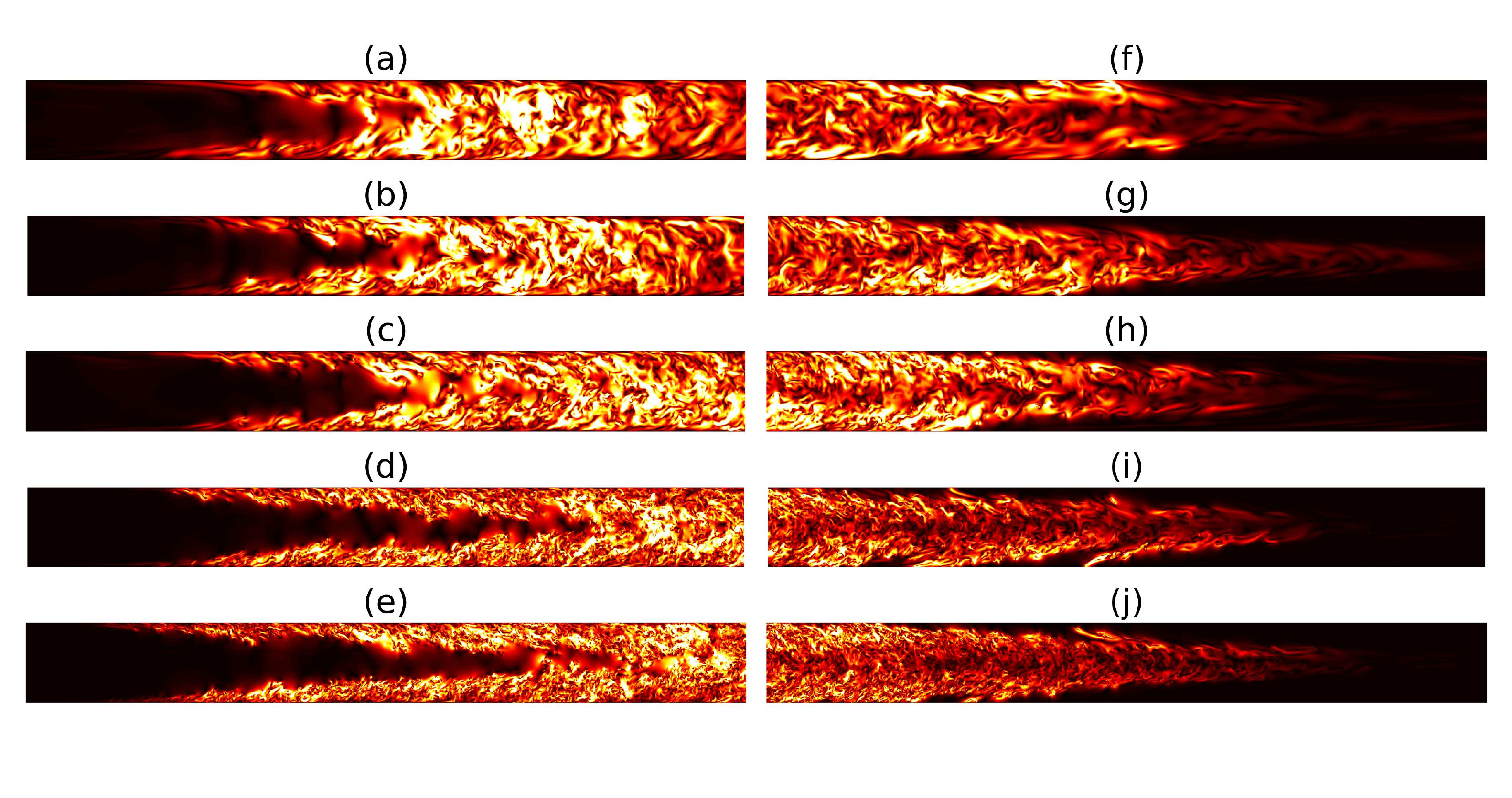}
\caption{\label{fig:metamorphosis} Fronts as $Re$ increases. Left column shows UF (panels a-e) and right column shows DF (panels f-j). The Reynolds numbers, from top, are 5000, 7500, 10000, 25000 and 40000. A $9D$-long pipe section is shown.}
\end{figure}

With the parameter settings for the $17.5D$ pipe as shown in table \ref{tab:parameter_setting_17.5D}, we simulated the fronts and measured the front speed up to $Re=40000$. The structure of the fronts at several Reynolds numbers ranging from 5000 to 40000 are visualised in figure \ref{fig:metamorphosis}. Overall, as $\Rey$ increases, the front regions on both the upstream and downstream become longer in the axial direction, i.e. visually the fronts become more slender and reach farther into the laminar region. The measured speeds are shown in table \ref{tab:parameter_setting_17.5D}.
\begin{table}
\renewcommand\arraystretch{1.2}
\centering
\begin{tabular}{p{0.5cm}p{1cm}p{0.7cm}p{0.7cm}p{0.7cm}p{0.7cm}p{0.7cm}p{0.7cm}p{1.5cm}p{1.7cm}p{0.7cm}p{0.7cm}}
& $\Rey$ & $L$	& $z_{f0}$	& $z_0$	&  $R $	& $A$	& $B$	& threshold  & \text{max}$\Delta t$  & $T$ & speed \\
\hline
\multirow{6}{*}{\rotatebox[origin=c]{90}{DF}}  
& 7500	&17.5 	& 5	& 15	& 2.25	& 0.30	& 0.25	& $5\times 10^{-4}$ & $1.25\times 10^{-3}$ & 1000	& 1.650\\
& 10000	&17.5	& 6	& 15	& 2.25	& 0.35	& 0.25	& $5\times 10^{-4}$ & $1.25\times 10^{-3}$ & 1350	& 1.710\\
& 17500	&17.5	& 5	& 15	& 2.25	& 0.35	& 0.25	& $5\times 10^{-4}$ & $1.00\times 10^{-3}$ & 390 	& 1.793\\
& 25000 &17.5	& 5	& 15	& 2.25	& 0.35  & 0.25	& $5\times 10^{-4}$ & $5.00\times 10^{-4}$ & 180	& 1.867\\
& 25000	&17.5	& 9	& 15	& 2.25	& 0.35	& 0.25	& $5\times 10^{-4}$ & $6.25\times 10^{-4}$ & 180	& 1.847\\
& 40000	&17.5	& 9	& 15	& 2.25	& 0.35	& 0.25	& $5\times 10^{-4}$ & $3.00\times 10^{-4}$ & 100	& 1.877\\
\hline
\multirow{5}{*}{\rotatebox[origin=c]{90}{UF}} 
& 7500	&17.5	& 10.5	& 4.75	& 4.25	& 0.30	& 0.25	& $5\times 10^{-4}$ & $1.25\times 10^{-3}$ & 750	& 0.511\\
& 10000	&17.5	& 10.5	& 4.75	& 4.25	& 0.30	& 0.25	& $5\times 10^{-4}$ & $1.25\times 10^{-3}$ & 500	& 0.447\\
& 17500	&17.5	& 10.5	& 5.5	& 2.5	& 0.35	& 0.25	& $5\times 10^{-4}$ & $5.00\times 10^{-4}$ & 100 	& 0.343\\
& 25000	&17.5	& 10.5	& 5.5	& 2.5	& 0.35	& 0.25	& $5\times 10^{-4}$ & $3.10\times 10^{-4}$ & 100	& 0.290\\
& 40000	&17.5	& 10.5	& 5.5	& 2.5	& 0.35	& 0.25	& $5\times 10^{-4}$ & $1.25\times 10^{-4}$ & 25		& 0.233\\
\end{tabular}
\caption{\label{tab:parameter_setting_17.5D} Reynolds number, damping parameters, threshold in $q$, time-step size and averaging time of the speed for the DF (top) and UF (bottom) in the $17.5D$ pipe. All parameter settings assure that $q$ drops by at least four orders of magnitude as turbulence passes the damping region and that $q$ drops away from the front naturally by approximately four orders of magnitude (see the example for $Re=5000$ in figure \ref{fig:crosssec0180_Re5000}). Two settings of $z_{f0}$ are compared for the DF at $\Rey=25000$, which differ by approximately $1\%$ on the front speed.}
\end{table}

\begin{figure}
\centering
\includegraphics[width=0.99\linewidth]{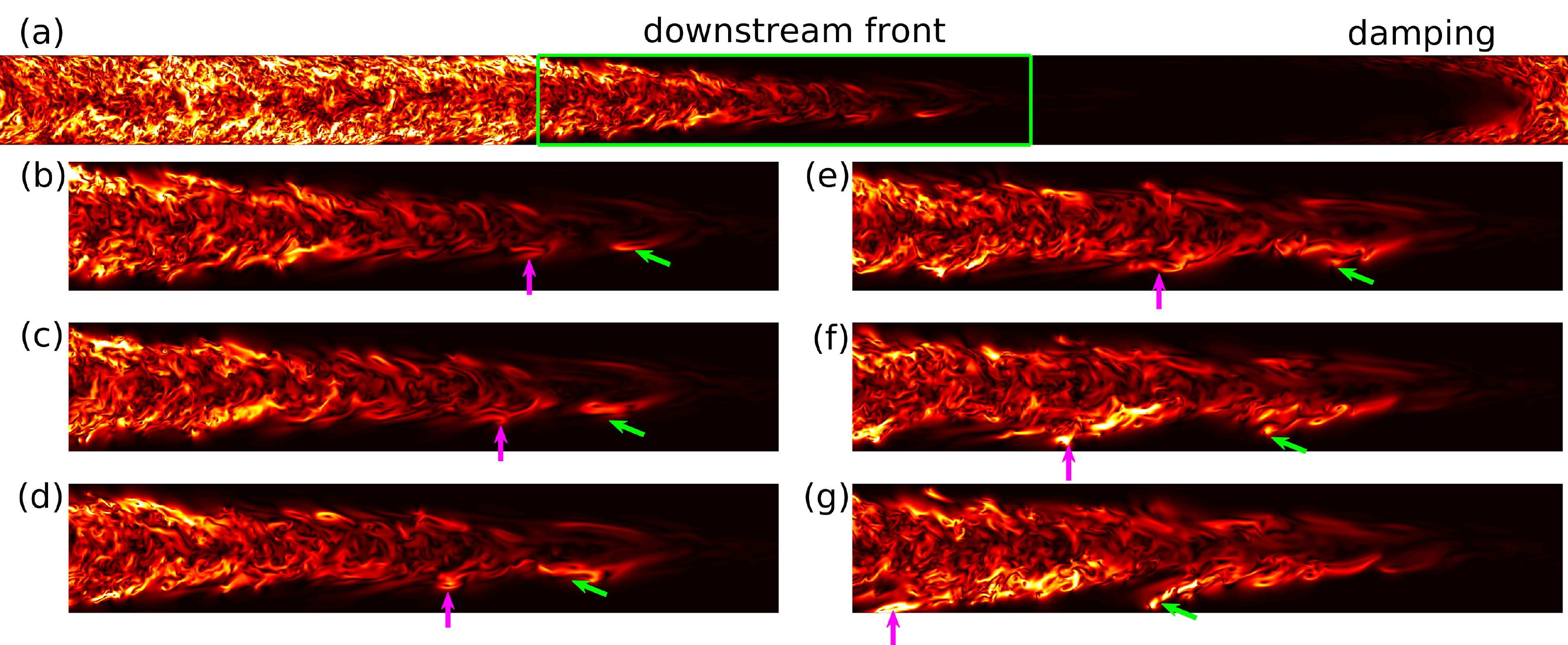}
\caption{\label{fig:Re25000_four_snapshots} Visualisation of local transition at the DF at $Re=25000$ in the moving frame of reference. (a) A snapshot of the cross-stream velocity $\sqrt{u_r^2+u_\theta^2}$ in the $z-r$ plane of the $17.5D$ pipe. Panels (b-g) show the flow in the region enclosed by the green rectangle in panel (a) at several time instants. Consecutive panels are separated by 0.875$D/U$. In panels (b-g), vertical arrows and tilted arrows mark two local transition events.}
\end{figure}

\begin{figure}
\centering
\includegraphics[width=0.99\linewidth]{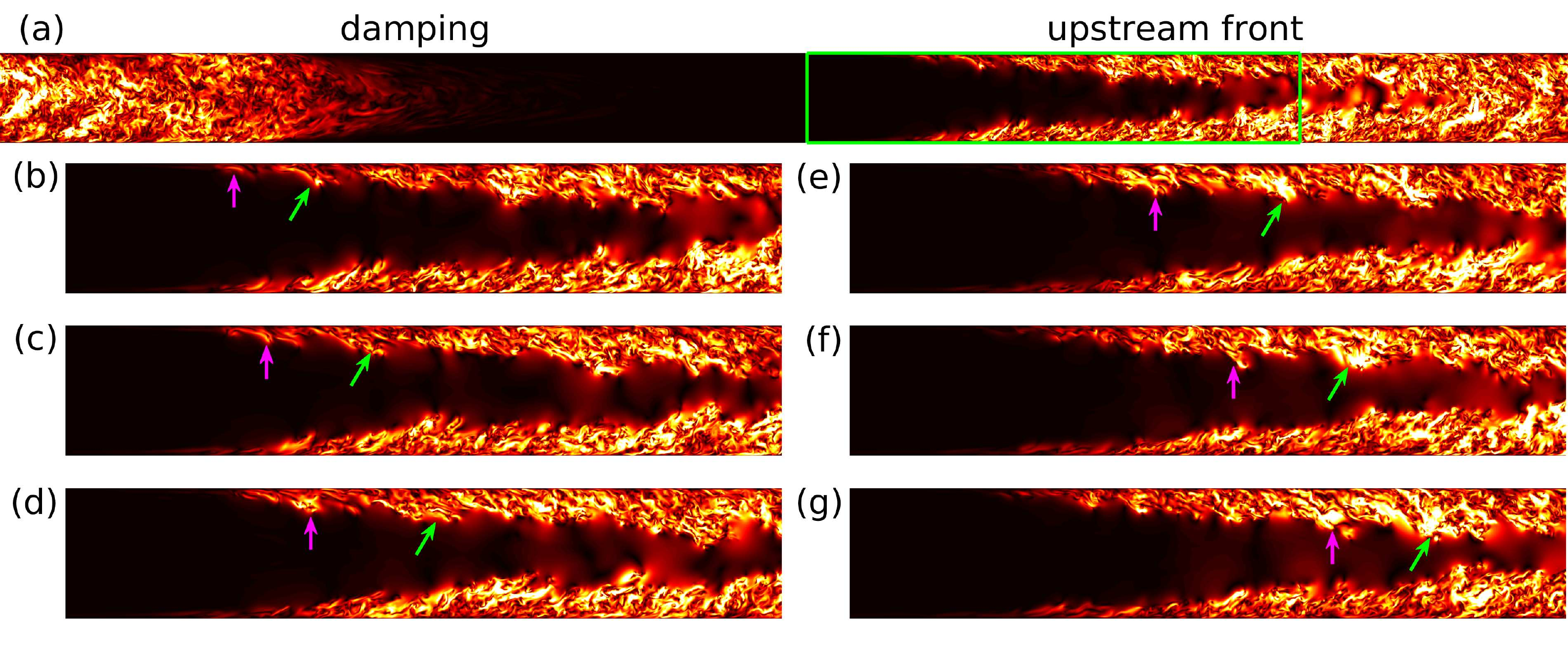}
\caption{\label{fig:Re25000_six_snapshots_UF} Visualisation of local transition at the UF at $Re=25000$ in the moving frame of reference. The same quantity as in figure \ref{fig:Re25000_four_snapshots} is plotted. In panels (b-g), consecutive panels are separated by 0.625$D/U$.}
\end{figure}

However, at $Re=40000$ the number of grid points becomes very large even for the $17.5D$ pipe, which reaches approximately $8.5\times 10^8$. The other restricting factor is the fastly decreasing time-step size. Our time-step size controller gives approximately $1.25\times 10^{-4}D/U$ to assure convergence for simulating the UF and $2.5\times 10^{-4} D/U$ for simulating the DF, given the same grid resolution setting, due to the explicit treatment of the damping term in the time-stepping. Therefore, the computational cost becomes so high that we can only afford to measure the DF speed over a time window of $100D/U$ and the UF speed over a time window of $25D/U$. In order to obtain reliable statistics of the front speed at $Re=40000$ and to consider higher $\Rey$, we had to reduce the pipe length further to reduce the cost. 
 
The following observation suggests that further reduction in the pipe length is possible. Figure \ref{fig:Re25000_four_snapshots}(a) visualises the DF at $\Rey=25000$ simulated in a $17.5D$ pipe. The flow inside the green rectangle enclosing a part of the DF is visualised at a few time instants in figure \ref{fig:Re25000_four_snapshots}(b-g). Local transition to turbulence at the tip of the front (close to the pipe center) can be observed, see the evolution of the flow structures pointed to by the vertical and tilted arrows in figure \ref{fig:Re25000_four_snapshots}(b-g). Initiating near the pipe center, these structures appear to be stretched and strengthen while extending toward the high shear region near the pipe wall. The slowing down (moving to the left) of the structures while approaching the wall, mainly due to the decreasing local flow speed, is clearly indicated by the axial locations of their upstream tips (see the arrows) in this frame of reference co-moving with the DF. In fact, because of the radial turbulent momentum transport, the generated turbulent fluctuations will also extend toward the pipe center and be advected downstream by the high speed flow at the pipe center. This possibly triggers successive local transitions at the tip of the DF. However, as the flow is already turbulent near the pipe center, this feedback process cannot be clearly seen. See supplementary movie 1 for more detailed dynamics at the DF. At the UF, similar local self-sustaining scenario was observed as shown in figure \ref{fig:Re25000_six_snapshots_UF}, but the difference is that the local transition initiates in the near wall region. The generated turbulent fluctuations feed back the near wall region and also \xu{extend} toward the pipe center and \xu{speed} up due to the increasing local flow speed. The turbulence eventually merges at the pipe center, filling the whole pipe cross-section. See supplementary movie 2 for more details. (We will revisit the transition scenario at the front tips in section \ref{sec:trend_of_speed}). {\color{black} \citet{Shimizu2009} and \citet{Duguet2010b} reported similar processes at the UF of puffs and slugs at much lower Reynolds numbers.} According to this scenario, we speculate that, the transition process is locally self-sustained at the front tips and does not depend on the flow sufficiently far away. This localness makes further pipe-length reduction possible.

\begin{figure}
\centering
\includegraphics[width=0.99\linewidth]{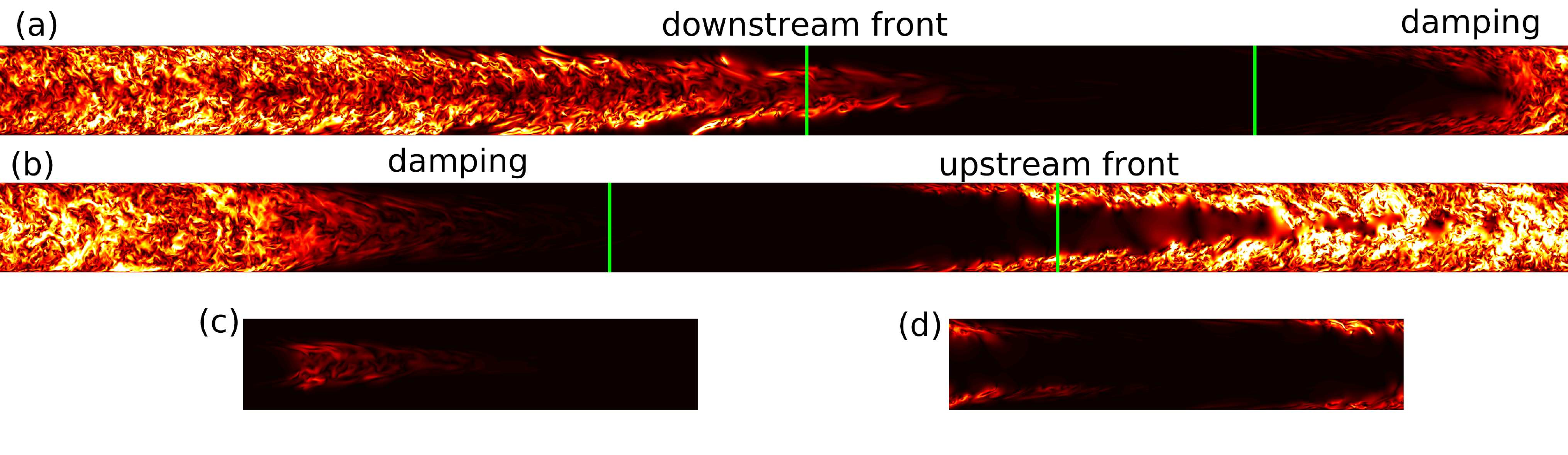}
\caption{\label{fig:Re25000_17.5D_5D} The $L=17.5D$ pipe (a, b) and the $5D$ pipe (c, d) at $Re=25000$. Panel (a) shows the DF and panel (b) shows the UF. The vertical green lines in each panel enclose the part of the front that we isolate by using the $5D$ periodic pipe, as shown in panel (c) and (d), respectively. The colourmap shows transverse velocity $\sqrt{u_r^2+u_\theta^2}$ in the $r$-$z$ cross-section. The specific damping parameters are listed in table \ref{tab:parameter_setting_17.5D} ($17.5D$ pipe) and table \ref{tab:parameter_setting_5D} ($5D$ pipe). Supplementary movie 3 and 4 show the flow in the $5D$ pipe at the DF and UF, respectively.}
\end{figure}

\begin{table}
\renewcommand\arraystretch{1.2}
\centering
\begin{tabular}{p{0.5cm}p{1cm}p{0.5cm}p{0.7cm}p{1.0cm}p{1.0cm}p{0.7cm}p{0.7cm}p{1.5cm}p{1.7cm}p{0.7cm}p{0.7cm}}
& $\Rey$  & $L$	& $z_{f0}$	& $z_0$	&  $R $	& $A$	& $B$	& threshold  & $\text{max}\Delta t$  & $T$ & speed \\
\hline
\multirow{6}{*}{\rotatebox[origin=c]{90}{DF}}  
& 17500	&   5 & 1.75	& 4.375	& 1.125	& 0.4	& 0.1	& $1\times 10^{-4}$ & $1.0\times 10^{-3}$ & 390 & 1.808\\
& 25000	&   5 & 1.25	& 4.375	& 1.125	& 0.4	& 0.1	& $1\times 10^{-4}$ & $7.5\times 10^{-4}$ & 400	& 1.853\\
& 40000	&   5 & 1.75	& 4.375	& 1.125	& 0.4	& 0.1	& $1\times 10^{-4}$ & $7.5\times 10^{-4}$ & 250	& 1.893\\
& 40000 &   5 & 1.25  	& 4.375	& 1.125 & 0.5   & 0.1   & $1\times 10^{-4}$ & $7.5\times 10^{-4}$ & 100	& 1.890\\
& 60000 &   5 & 1.75	& 4.375 & 1.125 & 0.4	& 0.1 	& $1\times 10^{-4}$ & $4.0\times 10^{-4}$ & 150	& 1.913\\
& 100000 &  5 & 1.25	& 4.375 & 1.125 & 0.4	& 0.1 	& $1\times 10^{-4}$ & $3.0\times 10^{-4}$ & 100	& 1.930\\
\hline
\multirow{4}{*}{\rotatebox[origin=c]{90}{UF}}  
& 17500	& 5 & 3.5	& 1.375	& 1.375	& 0.5	& 0.1	& $1\times 10^{-4}$ & $5.0\times 10^{-4}$ & 200 & 0.332\\
& 25000	& 5 & 3.5	& 1.375	& 1.375	& 0.5	& 0.1	& $1\times 10^{-4}$ & $5.0\times 10^{-4}$ & 200 & 0.287\\
& 40000	& 5 &3.5	& 1.375	& 1.375	& 0.5	& 0.1	& $1\times 10^{-4}$ & $3.75\times 10^{-4}$ &100 & 0.225\\
& 60000 & 5 & 3.5	& 1.375	& 1.375	& 0.5	& 0.1 	& $1\times 10^{-4}$ & $2.5\times 10^{-4}$  & 55	& 0.187\\
\end{tabular}
\caption{\label{tab:parameter_setting_5D} The Reynolds number, damping parameters, threshold in $q$, time-step size and averaging time of the speed for the DF (top) and UF (bottom) in the $5D$ pipe. Two settings for $z_{f0}$ are compared for the DF at $Re=40000$.}
\end{table}

We reduced the pipe length further to $L=5D$ to isolate the front tip, see the illustration in figure \ref{fig:Re25000_17.5D_5D} for $\Rey=25000$ (see also supplementary movie 3 and 4). This also relieves the restriction on the grid spacing and time-step size because the flow is relatively less turbulent at the very tip of the fronts, see Appendix \ref{sec:appendix}. The front speeds are measured for $Re=17500$ and $Re=25000$ again for validation, see table \ref{tab:parameter_setting_5D}. At $\Rey=17500$, the averaging time for the downstream front is the same as in the $17.5D$ pipe, whereas at $\Rey=25000$ the averaging times are significantly enlarged (see table \ref{tab:parameter_setting_17.5D} and \ref{tab:parameter_setting_5D}). For these two Reynolds numbers, we obtained very close speeds in the two pipes for both the UF and DF, justifying the use of the $5D$ pipe at least for the speed measurement (see table \ref{tab:parameter_setting_17.5D} and \ref{tab:parameter_setting_5D}). We were able to significantly increase the averaging time of the front speeds for $\Rey=40000$, and surprisingly, obtained very close values to those averaged over much shorter times in the $17.5D$ pipe. This suggests that the saturated front speeds at high $\Rey$ do not fluctuate significantly with time, as \citet{Song2014} reported. We then further measured the DF speed at $\Rey=60000$ and $10^5$ and the UF speed at $Re=60000$, see table \ref{tab:parameter_setting_5D}.

\subsection{The scaling of the front speed at high Reynolds numbers}

\begin{figure}
\centering
\includegraphics[width=0.99\linewidth]{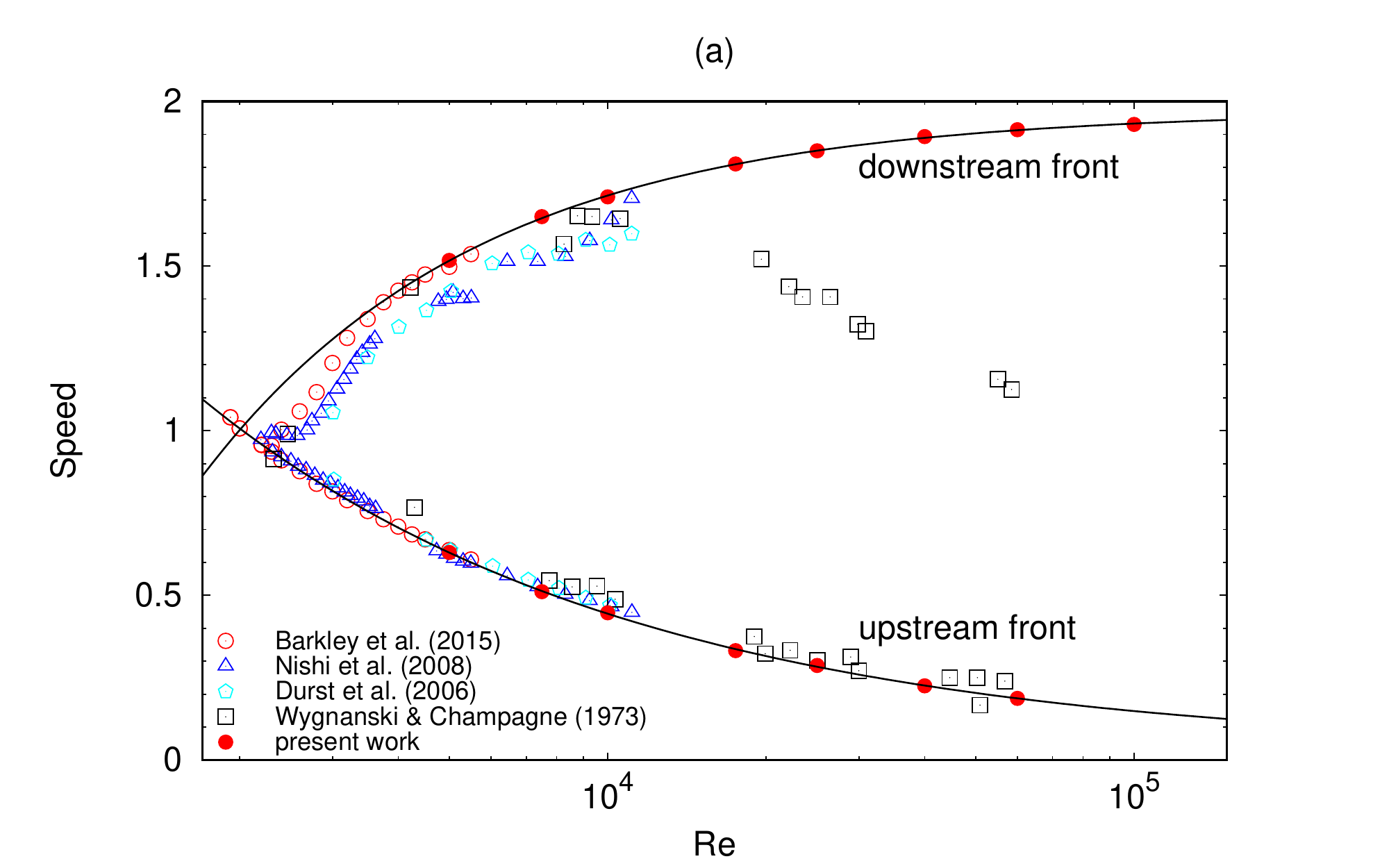}
\includegraphics[width=0.99\linewidth]{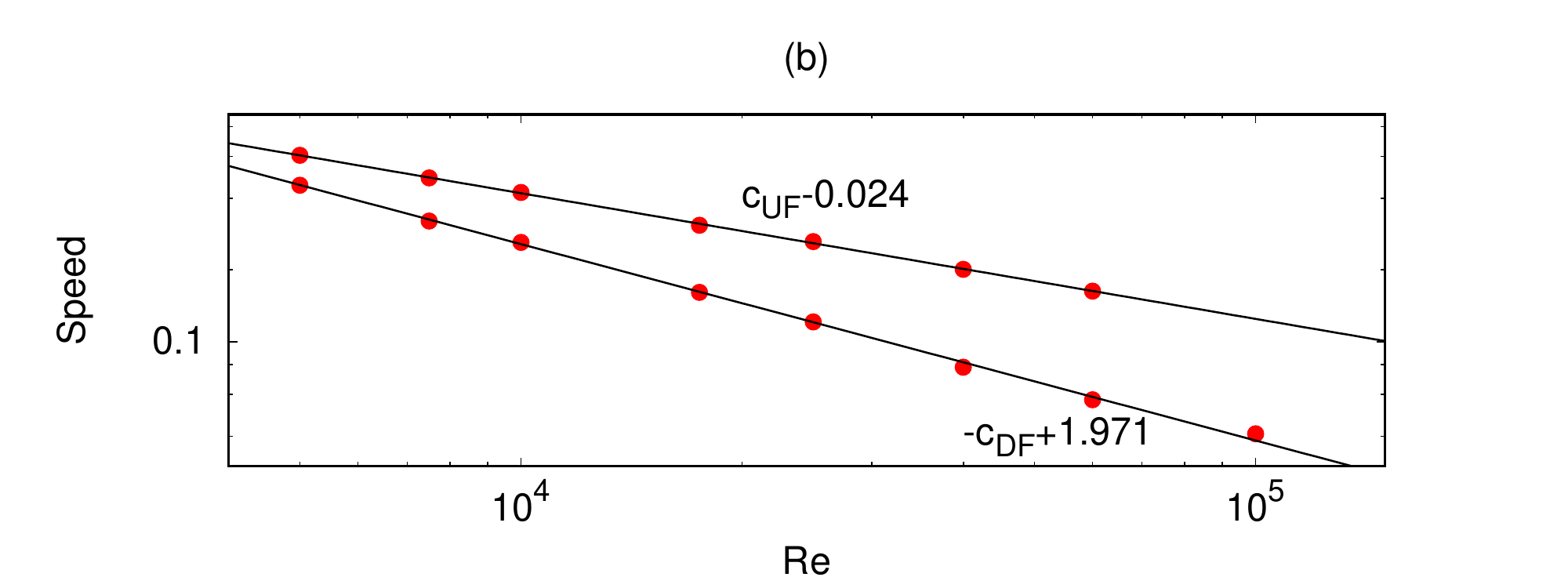}
\caption{\label{fig:front_speed_scaling} (a) Front speed as a function of $\Rey$. Some data sets from figure \ref{fig:front_speed_literature} are repeated here for comparison. (b) Our data (symbols) plotted in log-log scale. Lines show the fits (\ref{equ:scaling_DF}) and (\ref{equ:scaling_UF}).}
\end{figure}

Figure \ref{fig:front_speed_scaling} concludes the front speeds we measured with a comparison with most relevant data from the literature, which are also shown in figure \ref{fig:front_speed_literature}. The major difference with the former data sets is that our results give an increasing trend for the DF speed up to $\Rey=10^5$, whereas \citet{Wygnanski1973} gave the opposite trend above $\Rey\simeq 10000$. Besides, our data \xu{exhibit} less scattering.

Intuitively, the DF is not expected to propagate faster than the centerline velocity of the basic laminar flow, and also, the UF is not expected to propagate upstream in the laboratory frame of reference against the advection by the basic flow. Based on our results, we expect both speeds to keep monotonic with $\Rey$ and asymptotically approach some values in the range of $[0, 2]$ as $\Rey$ increases. We seek for a scaling of the form of $a+b\Rey^{c}$ for both front speeds, suggested by the model analysis of \citet{Barkley2015}. With the data for $\Rey=5000$ in table \ref{tab:parameter_setting}, for $\Rey=7500$ and 10000 in table \ref{tab:parameter_setting_17.5D} and for $\Rey=17500$ to $10^5$ in table \ref{tab:parameter_setting_5D}, we obtained best fits with the least square errors and slightly reformulated the scalings as
\begin{align}
c_{\text{DF}}&\approx 1.971-(Re/1925)^{-0.825}\label{equ:scaling_DF},\\
c_{\text{UF}}&\approx 0.024+(Re/1936)^{-0.528}\label{equ:scaling_UF},
\end{align} 
where $c_{\text{DF}}$ denotes the DF speed and $c_{\text{UF}}$ the UF speed. These two scalings are plotted as black solid curves in figure \ref{fig:front_speed_scaling}(b), which can be seen \xu{to} fit our data (red filled circles) very well. Besides, the scaling of the $c_{\text{UF}}$ (\ref{equ:scaling_UF}) also fits very well the data from the literature. One can notice that the scaling of $c_{\text{DF}}$ also fits well the data from \citet{Barkley2015} down to $\Rey\simeq 3500$, while deviations are observed at lower $\Rey$. \BS{This can be expected because the fit (\ref{equ:scaling_DF}) is based on data for strong fronts at high $Re$, while the DF is weak below $\Rey\simeq 2900$ and the transition from weak DF to strong DF completes close to $\Rey\simeq 3500$ according to \citet{Song2017}.}

\section{Discussion}
\label{sec:discussion}
 
\citet{Barkley2015} by theoretical modelling and asymptotic analysis in a model system proposed that, in the limit of asymptotically strong front, the speeds of both fronts obey the scaling of $a+b\Rey^{-0.5}$, where $a$ and $b$ are constants and are different for the UF and DF. Their DNS and experimental data up to $\Rey=5500$ in long pipes showed that the UF speed seems to approximately follow  this scaling. Their data for the DF in the strong front regime (above $\Rey\simeq 3500$) \xu{seem} to follow this scaling also. However, the $\Rey$ range for assessing the scaling is too narrow (from 3500 to 5500). Here, our simulations up to $\Rey=60000$ for the UF show that the UF speed indeed closely follows the scaling proposed by \citet{Barkley2015} \BS{based on data in a much smaller $\Rey$ range}, but the DF speed approximately follows (\ref{equ:scaling_DF}) from $\Rey=5000$ to $10^5$, in which the power considerably deviates from $-0.5$. 
This disagreement suggests that their assumption that the UF and DF become the mirror image of each other at high $\Rey$ may not hold in real pipe flow. 

Undoubtedly, results in the $\Rey$ regime of the present work may not be simply extrapolated to infinite $\Rey$, and we are not proposing the scalings (\ref{equ:scaling_DF}) and (\ref{equ:scaling_UF}) as the asymptotic scalings as $\Rey\to\infty$. Besides, our results for high $\Rey$ are based on measurements over $\mathcal{O}(100)$ time units. Although the results of \citet{Song2014} showed that the fluctuation of the front speed decreases as $\Rey$ increases and therefore the front speed measured in a reasonably long time interval will be increasingly representative as $\Rey$ increases, we cannot rule out the possibility of considerable migration over larger time spans in the front speed (especially for the DF). Nonetheless, in the following, we propose that the monotonicity of the characteristic speed of both fronts will persist as $\Rey$ increases further.

\subsection{Production and dissipation of kinetic energy at the fronts}
As we showed in figure \ref{fig:Re25000_four_snapshots} and figure \ref{fig:Re25000_six_snapshots_UF}, the local transition at the DF is initiated near the pipe center and close to the pipe wall at the UF. This observation can be more quantitatively shown by the energy budget analysis. 

Following \citet{Song2017}, we calculated the production and dissipation of the kinetic energy \xu{associated} with velocity fluctuations about the mean flow in the frame of reference co-moving with the fronts, and the calculation is performed for the two fronts separately. Specifically, the production $P$ and dissipation $\epsilon$ are calculated as
\begin{equation}\label{equ:P_E_terms}
P=-\overline{ u_i' u_j'} \frac{\partial \overline {u_i}}{\partial x_j},~~~~~~ \epsilon=\frac{2}{Re}\overline{ s_{ij}s_{ij}},
\end{equation}
in which the overbar denotes the average over time and over the azimuthal direction in the moving frame of reference, the prime denotes the velocity fluctuation with \xu{respect} to the mean flow $\bar{\boldsymbol u}(r,z)$, $x_j$ denotes the spatial coordinates and $s_{ij}$ is the fluctuating rate of strain defined as
\begin{equation}
s_{ij}=\frac{1}{2}\left( \frac{\partial u_i'}{\partial x_j}+\frac{\partial u_j'}{\partial x_i}\right).
\end{equation}

\begin{figure}
\centering
\includegraphics[width=0.99\linewidth]{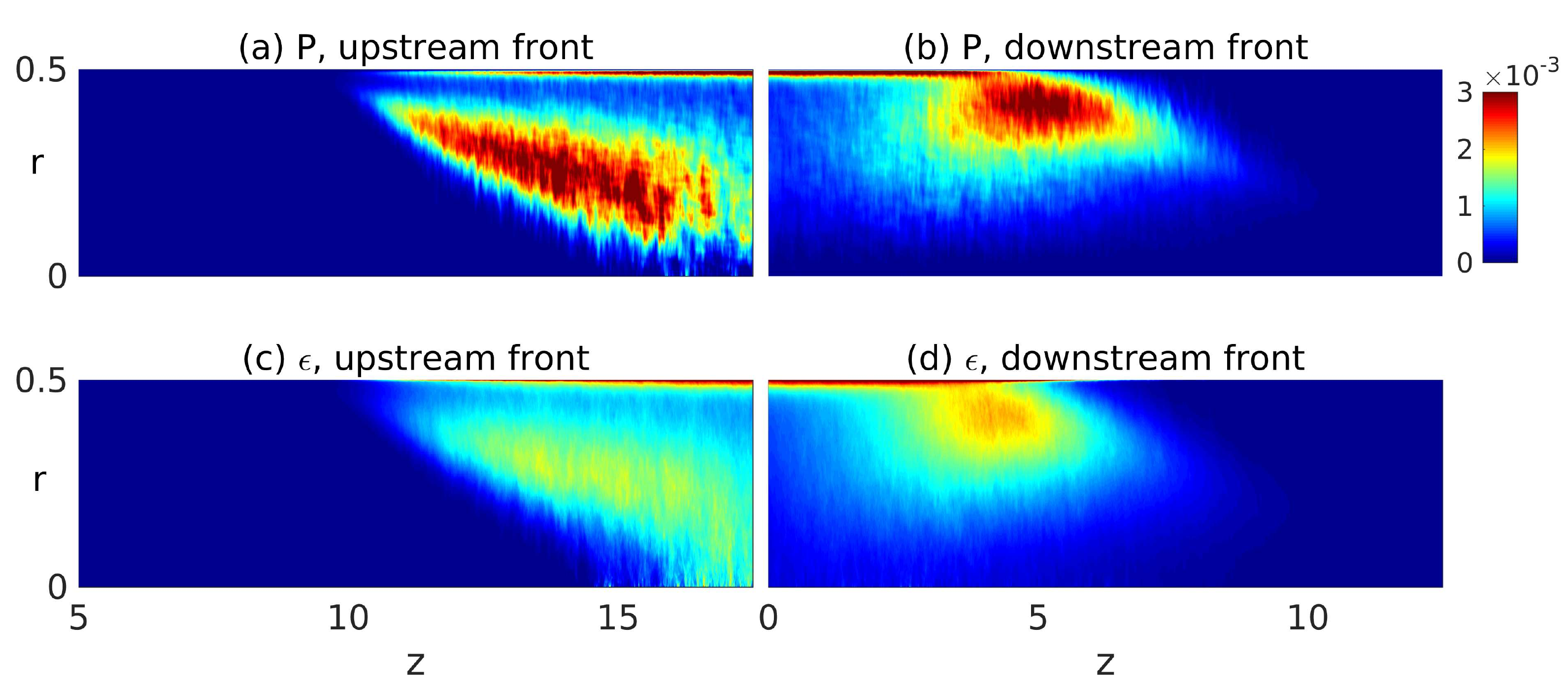}
\caption{\label{fig:energy_budget_Re25000} Azimuthally averaged production and dissipation in the $r-z$ cross-section at the UF (a, c) and DF (b, d) {\color{black} for $Re=25000$}. The length in the vertical direction is stretched by a factor of 8 in order to better display the distribution of the terms in the radial direction. In all panels, a pipe segment of $12.5D$ is shown.}
\end{figure}

Figure \ref{fig:energy_budget_Re25000} shows our calculation of $P$ and $\epsilon$ at the fronts for $Re=25000$ in the $17.5D$ pipe. It can be seen that at the UF (panel a), production is highest in a thin layer near the pipe wall, as in the fully developed bulk region \citep{Dimitropoulos2001,Schlatter2013}, and in a more extended tilted region stretching from the near wall region to the pipe center, which extends more than seven diameters long in the axial direction. There is a gap with relatively lower production rate between these two regions. Similar distribution can be seen for the dissipation, except for that the dissipation rate in the extended tilted region is considerably lower than that in the near wall region. 
At the DF, the production and dissipation rates are highest in a tilted region significantly far from the wall, and the downstream tip of this region protrudes into the laminar flow region close to the pipe center, see figure \ref{fig:energy_budget_Re25000}(b).

\begin{figure}
\centering
\includegraphics[width=0.99\linewidth]{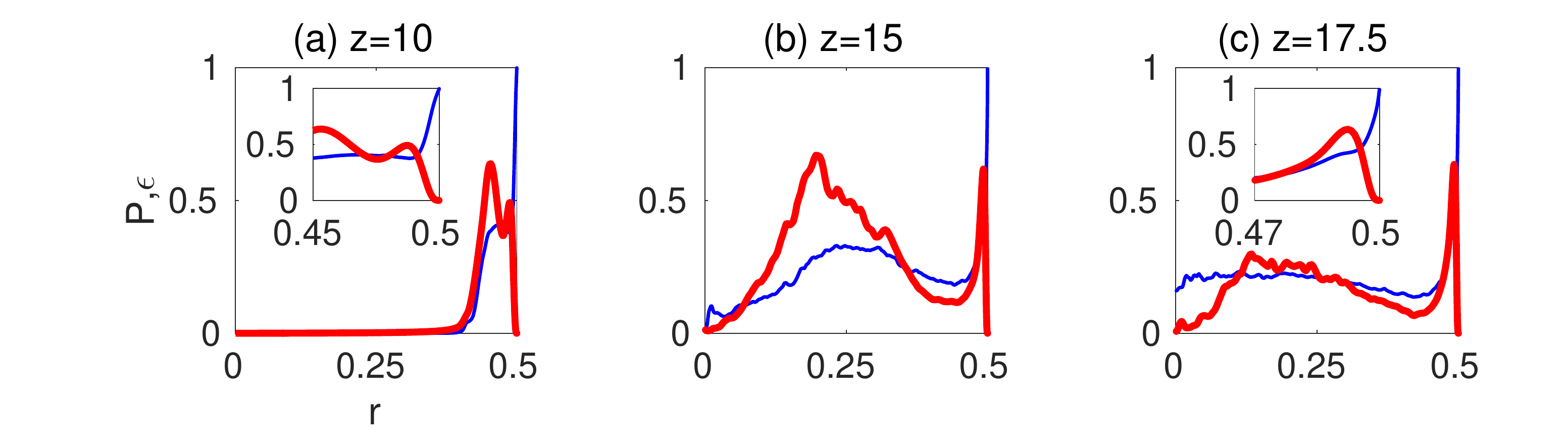}
\includegraphics[width=0.99\linewidth]{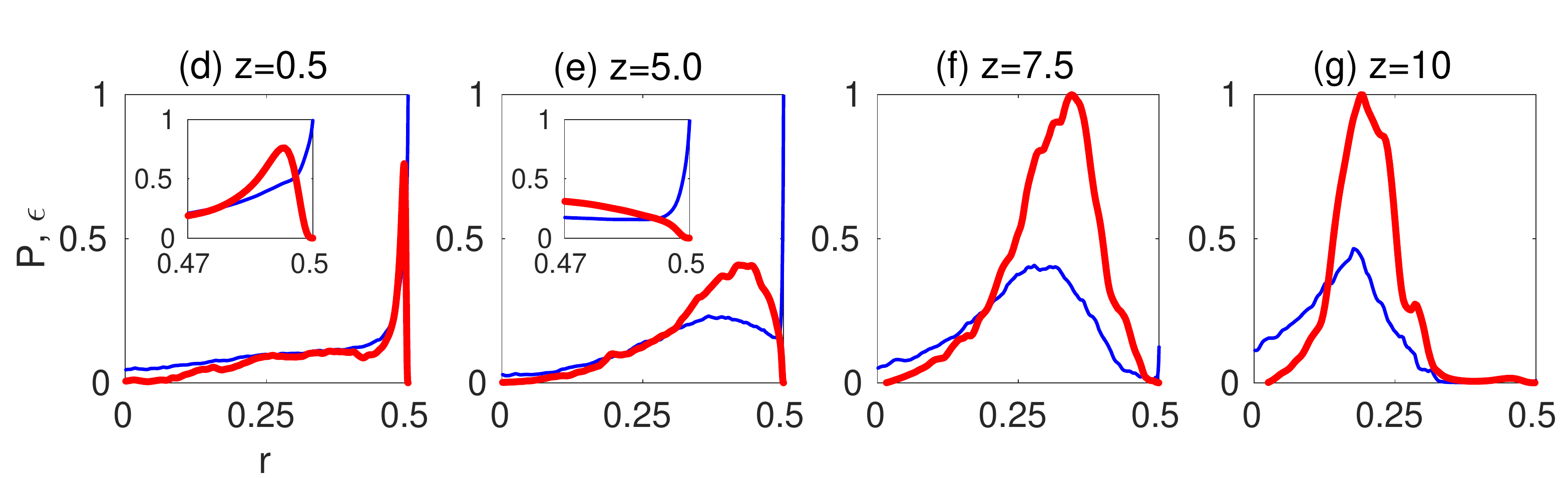}
\caption{\label{fig:P_E_profiles_Re25000} The profiles of production (bold red) and dissipation (thin blue) at several axial positions for the UF (panels a-c) and DF (panels d-g) {\color{black} for $Re=25000$}. The data are taken from that shown in figure \ref{fig:energy_budget_Re25000}. $P$ and $\epsilon$ are normalized by the maximum of the two over the radius, i.e. $\max\{\max_r{P},\max_r\epsilon\}$, at respective axial positions. The insets in panels (a, c, d, e) show the near wall region.}
\end{figure}

The radial distributions of $P$ and $\epsilon$ at the fronts can be more quantitatively shown by the radial profiles of the terms, see figure \ref{fig:P_E_profiles_Re25000} for $Re=25000$. The double-peak structure of $P$ and $\epsilon$ at the UF can be clearly seen in figure \ref{fig:P_E_profiles_Re25000}(b-c), which was also reported by \citet{Wygnanski1973}. Two peaks in $P$ can be observed even at the upstream tip of the front where $P$ and $\epsilon$ are very low, see figure \ref{fig:P_E_profiles_Re25000}(a). The two peaks correspond to the near wall region and the tilted region at the fronts as shown in figure \ref{fig:energy_budget_Re25000}(a, c), respectively. The flow at $z=0.5$ for the DF is close to a fully developed turbulent flow. In viscous unit, the peak of the production in figure \ref{fig:P_E_profiles_Re25000}(d) is approximately at $y^+\approx 10$, which is at the bottom of the buffer layer, whereas the dissipation is dominant in the sublayer and peaks at the wall. The production nearly vanishes close to the pipe center, whereas the dissipation is low but stays finite. These distributions of $P$ and $\epsilon$ are similar to typical profiles for fully developed turbulent flow from the literature \citep{Dimitropoulos2001,Schlatter2013}. Figure \ref{fig:P_E_profiles_Re25000}(e-g) shows that as the axial position moves toward the tip of the DF, the region for $P$ moves toward the pipe center and at $z=10$, the peak of $P$ is approximately at $r=0.2D$. Only a single peak can be observed at most part of the DF, in contrast to the UF. 

The distributions of $P$ and $\epsilon$ are consistent with our observation of the dynamics at the fronts illustrated by figures \ref{fig:Re25000_four_snapshots} and \ref{fig:Re25000_six_snapshots_UF}. At the UF, figure \ref{fig:Re25000_six_snapshots_UF} shows that the transition is initiated at the tip of the front, which is close to the pipe wall, and the generated flow structures extend toward the pipe center while being advected downstream by the faster flow. In this process, the turbulence strengthens presumably due to the stretching of the local mean shear (see evidence for the fronts of puffs by \citet{Holzner2013}). The turbulent kinetic energy will be produced during the spreading and strengthening of velocity fluctuations. Therefore, one can expect that a high-$P$ region would appear as a tilted region connecting the near wall region where the transition is initiated and the pipe center where turbulence merges and fills the whole pipe cross-section (see the tilted red region in figure \ref{fig:energy_budget_Re25000}a). With a strong production comes also a strong dissipation in the same region (see figure \ref{fig:energy_budget_Re25000}c), but the production outweighs the dissipation (see also figure \ref{fig:P_E_profiles_Re25000}b). The continuing transition and formation of turbulence at the tip (near pipe wall) and the spreading of the turbulence toward the pipe center together, counter-acting the distorting effect of the advection of the local mean flow, keep the characteristic shape of the front (see figure \ref{fig:metamorphosis}). At the DF, a similar transition scenario occurs as illustrated in figure \ref{fig:Re25000_four_snapshots}. The distinction to the UF is that the local transition at the front tip is initiated near the pipe center, and the generated turbulence spreads toward the pipe wall while being advected upstream relative to the front tip.

\subsection{The trend of the front speed as $\Rey\to\infty$}\label{sec:trend_of_speed}
Based on our observation of the dynamics at the fronts and the energy budget analysis, we propose that the trends of the front speeds will stay monotonic, i.e. the speed of the DF will keep increasing and the UF speed will keep decreasing as $\Rey$ increases further.

As the tip of the fronts can be self-sustained, the transition at the tip must be triggered by velocity disturbances locally. As the adjacent flow is laminar, the disturbances that trigger the transition necessarily originate from the turbulent region. One possibility is that, at the UF, velocity disturbances close to the pipe wall, which propagate at low speeds due to the slow advection by the local mean flow, protrude from the turbulent region and trigger the transition. The generated turbulence locally feeds back the near wall region with \xu{velocity} disturbances, closing the self-sustaining cycle. {\color{black} It was proposed in the literature for puffs at low Reynolds numbers that low speed streaks protrude from the turbulent region at the UF and cause instabilities (Kelvin-Helmholtz by \citet{Shimizu2009} and inflectional by \citet{Hof2010}), sustaining the puffs. Our observation suggests that similar mechanisms may also take part at much higher Reynolds numbers.} 
 At the DF, disturbances close to the pipe center, which propagate at high speed because of the advection of the mean flow, trigger the local transition at the front tip. Similarly, the generated turbulence feeds back the center region with velocity disturbances. The self-sustainment is the reason why the front tips can be isolated without significantly affecting the kinematics of the fronts. 

\begin{figure}
\centering
\includegraphics[width=0.99\linewidth]{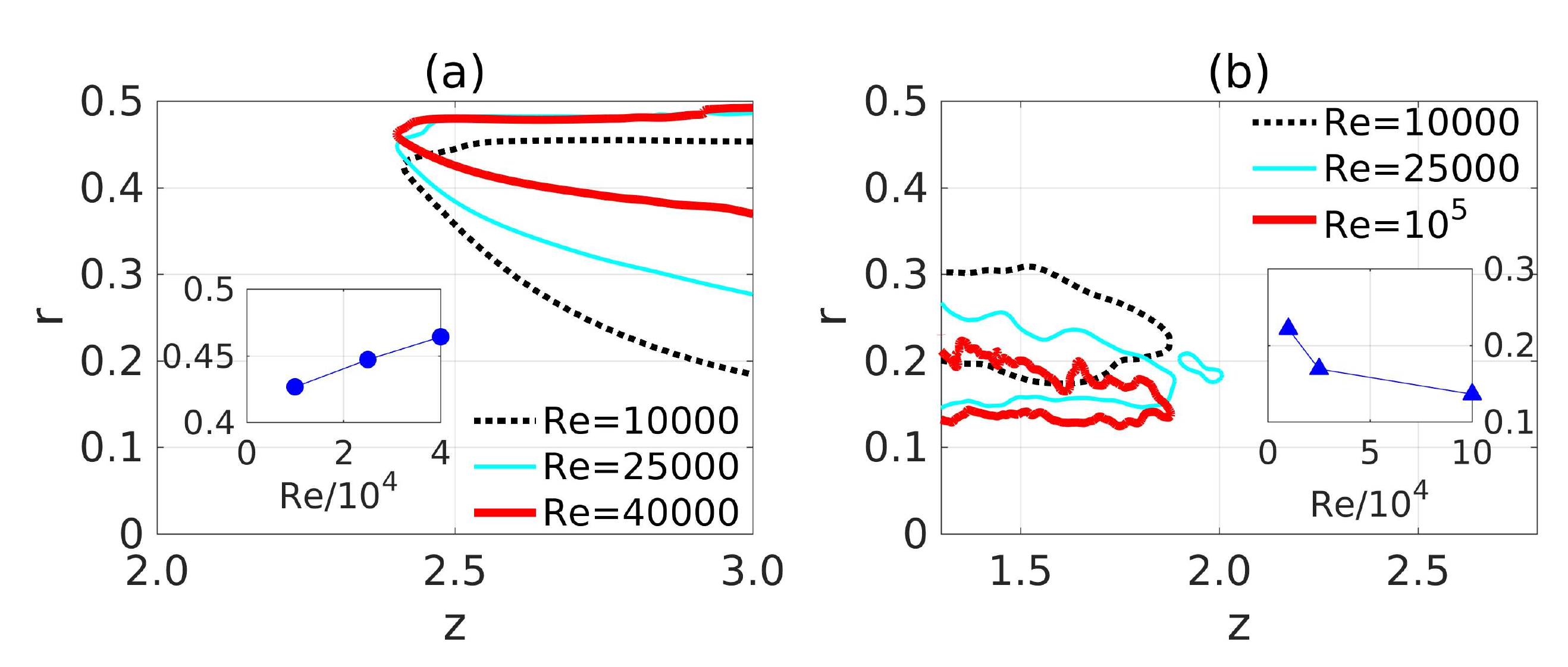}
\caption{\label{fig:contour_lines_at_fronts} Contour lines of the net production $P-\epsilon$ at the front tip plotted in the $r$-$z$ plane. (a) The contour level of $10^{-6}$ at the UF. (b) The contour level of $10^{-5}$ at the DF. The positions of the fronts are shifted in the axial direction such that the nose of these contour lines are approximately located at the same axial position for comparison. The ruggedness in some contour lines are due to the limited data for the energy budget analysis. The insets show the trend of the radial position of the nose of the contour lines, i.e. the leftmost point for the UF and the rightmost point for the DF.}
\end{figure}

Positive $P-\epsilon$, \xu{referred to as }the net production, is a signal for turbulence strengthening, therefore, presumably, can also be considered as a signature for the transition to turbulence at the front tip. It is expected that the net production would be very small at the early stage of the transition. Due to limited data for the energy budget analysis (we didn't save the velocity field frequently due to the large data size) and numerical errors, very low-level net production could be a false positive. Therefore, it is difficult to quantify the precise position of the transition at the front tip using this quantity in practice, because it is difficult to define a clear-cut threshold for the transition. Nevertheless,
the regions enclosed by the contour lines with low contour levels in figure \ref{fig:contour_lines_at_fronts} can still be used to illustrate the trend of the position of the local transition at the front tip as $\Rey$ increases. As can be seen, the region with net production moves closer to the pipe wall as $\Rey$ increases at the UF, whereas moves closer to the pipe center as $\Rey$ increases at the DF (especially, see the position of the noses of the contour lines shown in the insets of figure \ref{fig:contour_lines_at_fronts}). 
This observation indicates that, as $\Rey$ increases, the position of the front tip where transition is initiated, moves toward the pipe wall at the UF and moves toward the pipe center at the DF. In other words, transition-inducing velocity disturbances are located closer to the pipe wall at the UF tip and are located closer to the pipe center at the DF tip as $\Rey$ increases.

In the following, we propose that the propagation speed of transition-inducing disturbances at front tips should be largely determined by the local mean flow speed. Studies have shown that, at least in fully developed turbulent channel and pipe flows, the advection speed of velocity fluctuations is close to (slightly slower than) the local mean flow speed except for the region very close to the wall with $y^+\lesssim 10$, where the propagation speed of velocity fluctuations is considerably faster than the local mean flow \citep{Alamo2009,Pei2012,Wu2012}. Therefore, these studies suggest that the region where the propagation speed of velocity fluctuations significantly deviates from the local mean flow speed is the near-wall region where dissipation and production are large and strongly differ from each other (see the part of $r\gtrsim 0.48$ in figure \ref{fig:P_E_profiles_Re25000}d). While velocity fluctuations roughly follow the local mean flow in regions sufficiently far from the wall where production and dissipation are low or are nearly in balance, such as near the pipe center and at the front tips. Consequently, the propagation speed of the transition-inducing velocity disturbances at the front tips should be largely determined by the local mean flow speed. Therefore, the trends in the radial position of the front tips shown in figure \ref{fig:contour_lines_at_fronts} suggest that the UF speed should decrease and the DF speed should increase as $Re$ increases.
This argument is consistent with our data in the considered $\Rey$ range.

\begin{figure}
\centering
\includegraphics[width=0.99\linewidth]{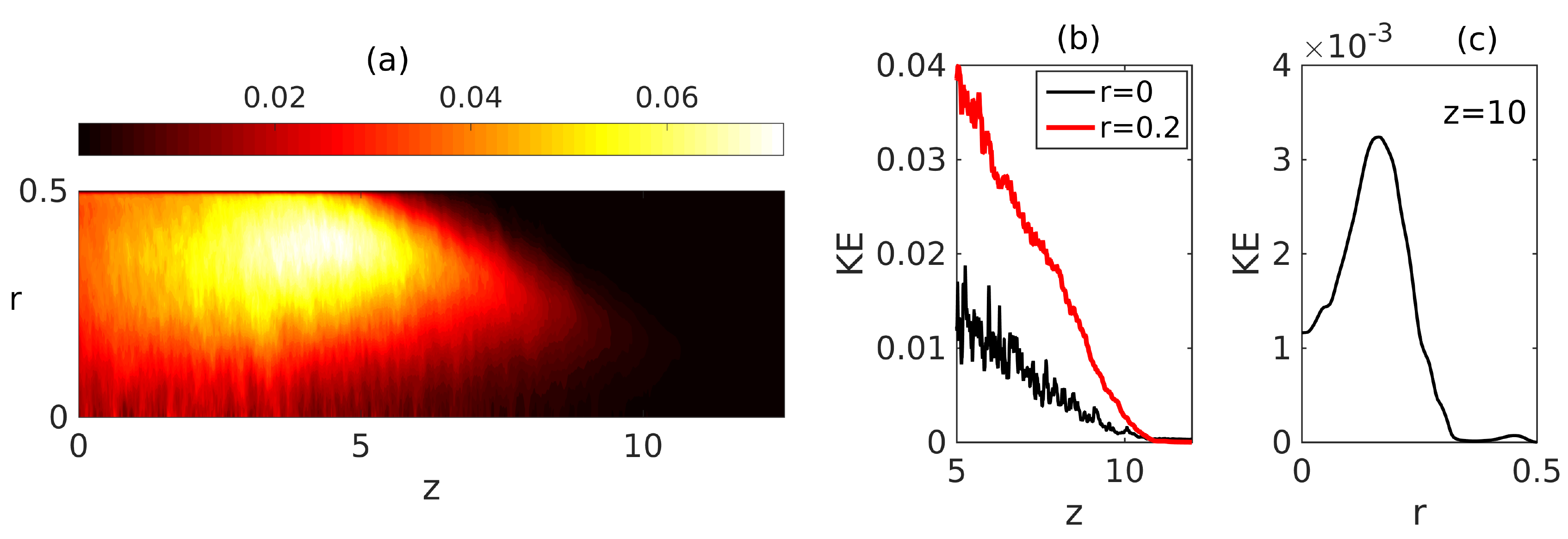}
\caption{\label{fig:KE_at_DF} (a) The distribution of the azimuthally and temporally averaged turbulent kinetic energy $KE=\overline{u_r'^2+u_\theta'^2+u_z'^2}$ in the $z-r$ cross-section at the DF. (b) The distribution of $KE$ along the pipe axis $r=0$ (thin black) and at $r=0.2$ (bold red). (c) The distribution of $KE$ along the radius at $z=10$.}
\end{figure}

Although we don't have data at further higher $\Rey$, the trends can be expected to persist as $\Rey$ increases further because of the following argument. It has been known that, as $\Rey$ increases, the amplitude of disturbances needed to trigger turbulence decreases \citep{Hof2003,Peixinho2007}. Therefore, it can be expected that the transition at the UF tip would occur closer to the pipe wall as $\Rey$ increases, because disturbances closer to the wall, which are in general of lower amplitudes, would be sufficient to trigger the transition. As a result, the speed of the UF would keep decreasing as $\Rey$ increases. At the DF tip, the closer to the pipe center the weaker the velocity disturbances (see figure \ref{fig:KE_at_DF} and the contours of $\sqrt{u_r^2+u_\theta^2}$ at the tip of the fronts in figure \ref{fig:Re25000_four_snapshots}), and the faster the disturbances propagate due to the faster advection by the local mean flow. 
This distribution at the DF can be understood from another perspective. \BS{As figure \ref{fig:P_E_profiles_Re25000} (e-g) show, the dissipation outweighs the production near the pipe center and the difference between the two terms is largest at the pipe center. Thus, the turbulent kinetic energy can be expected to be lower \xu{when} closer to the pipe center.} As weaker disturbances are sufficient to trigger transition as $\Rey$ increases, it can be expected that the position where transition initiates at the DF tip would move toward the pipe center as $\Rey$ increases, just as figure \ref{fig:contour_lines_at_fronts}(b) suggest. Consequently, it can be inferred that the DF speed will keep increasing as $\Rey$ increases. 

As far as we can see, there is no mechanism that causes the DF to decelerate toward the bulk speed in the limit of $\Rey\to\infty$. The decreasing trend \citet{Wygnanski1973} reported {\color{black} at high $\Rey$} should be attributed to the insufficient pipe length they used which did not {\color{black} even allow the basic laminar flow to fully develop before leaving the pipe exit, let alone the front speed. Their upstream front speed was not severely affected, because the developing blunted basic flow profile deviates from the parabolic profile most severely near the pipe center, but only slightly near the pipe wall given the cylindrical geometry (see their figure 9). In fact, the decreasing trend of their DF speed supports our argument that the front speed is largely determined by the local flow speed at the front tip, given that the basic flow is more blunted (less developed) in their pipe as $\Rey$ increases and consequently the local flow speed at the tip of the DF decreases.}

It is interesting to note that, if the fittings (\ref{equ:scaling_DF}) and (\ref{equ:scaling_UF}) would apply in the limit of $\Rey\to\infty$, one would see that the DF speed would not approach exactly $2U$ and that the UF speed would not approach zero precisely. This would imply that, even at infinite $Re$, pipe flow would not become absolutely unstable (UF speed stays finite), and that the transition-inducing velocity disturbances would propagate slightly slower than the centerline velocity of the basic laminar flow.

\subsection{The flow structures at the two front tips}
Although the mechanism of the local transition at the front is still far from being clear, based on our analysis, we speculate that the transition mechanism at the UF is possibly different from that at the DF, because the transition occurs in the high shear region near the wall at the UF but in the low shear region near pipe center at the DF. In the following, we show the flow structures at the two fronts.

\begin{figure}
\centering
\includegraphics[width=0.99\linewidth]{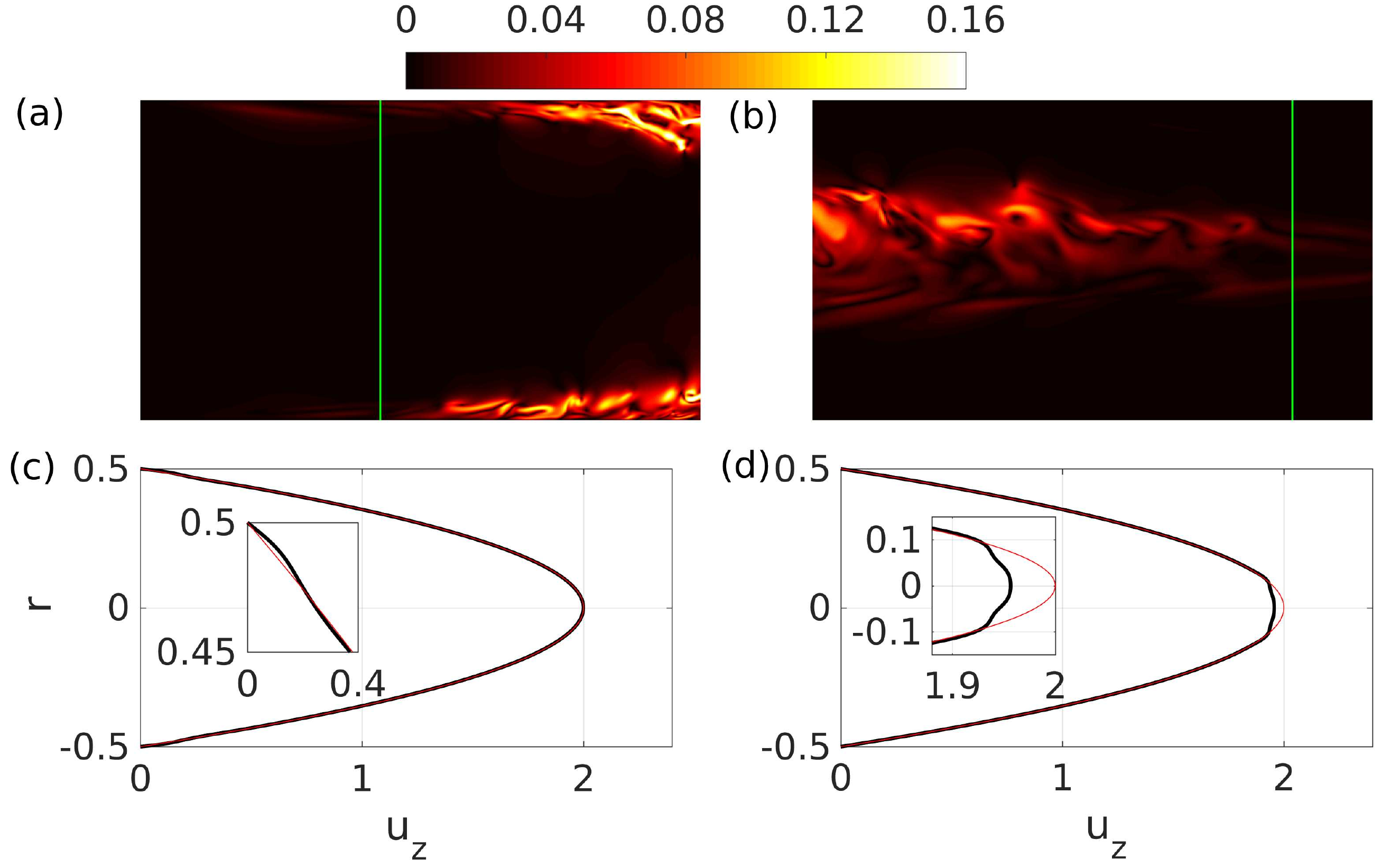}
\includegraphics[width=0.99\linewidth]{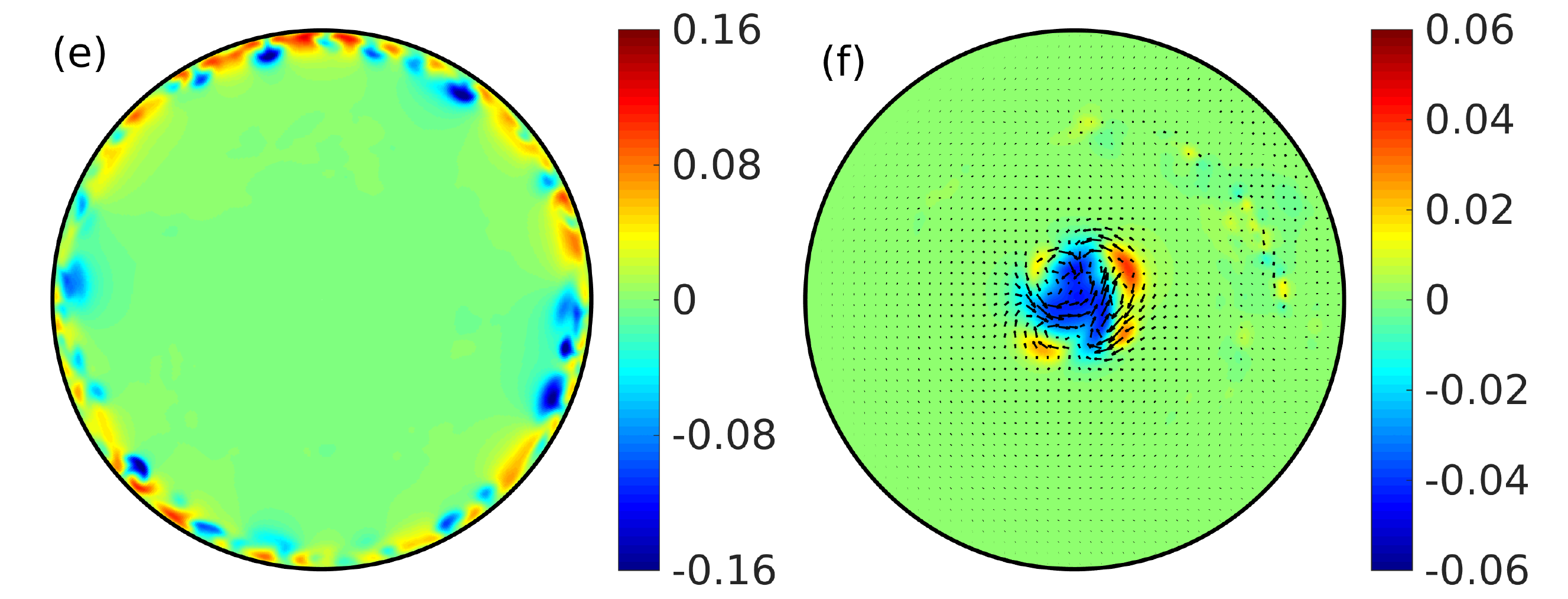}
\caption{\label{fig:flow_structure_Re40000} The cross-stream velocity $\sqrt{u_r^2+u_\theta^2}$ in the $z-r$ plane at the tip of the UF (a) and DF (b) at $\Rey=40000$. (c, d) The profiles of the streamwise velocity, averaged in the azimuthal direction, at the positions marked by the two vertical lines in panel (a) and (b), respectively. (e, f) The contours of streamwise velocity deviated from the parabola in the $r-\theta$ plane at the positions marked by the two vertical lines in panel (a) and (b), respectively. In (f), the in-plane velocity is also plotted as vectors.}
\end{figure}

Figure \ref{fig:flow_structure_Re40000} visualises an instantaneous flow field at the front tips at $Re=40000$ in the $5D$ pipe. The cross-stream velocity $\sqrt{u_r^2+u_\theta^2}$ near the front tips are plotted in \xu{figure}~\ref{fig:flow_structure_Re40000}(a, b). The shown region is the $z-r$ cross-section of a $1.7D$ pipe segment. The vertical lines mark the tip of the fronts, where, roughly, transition to turbulence is initiated (judged by eye). The azimuthally averaged streamwise velocity profiles, i.e. the local mean profiles, at the vertical lines are plotted in figure \ref{fig:flow_structure_Re40000}(c, d), which are nearly parabolic. Slight deviations can be observed very close to the wall at the UF and very close to the pipe center at the DF. In other words, the mean flow modification by the turbulence at the front tips is very weak. Figure \ref{fig:flow_structure_Re40000}(e, f) show the contours of streamwise velocity components in the $r-\theta$ cross-section at the two vertical lines in figure \ref{fig:flow_structure_Re40000}(a, b). At the UF tip, it can be seen that high- and low-speed streaks with high azimuthal wavenumbers prevail close to the pipe wall, whereas the flow is nearly laminar elsewhere. Although the cross-stream velocity is weak (see the contour plots in figure \ref{fig:flow_structure_Re40000}\xu{a}), the magnitude of the streaks can be as large as $\mathcal{O}(0.1)$, i.e. streaks are the dominant flow feature at the tip of the UF. At the DF tip, flow structures (streaks and vortices) are concentrated close to the pipe center. A low speed region dominates at the pipe center, but it can be seen that non-axisymmetric components are of substantial magnitudes even in the streamwise velocity component, unlike the weak DF at much lower $\Rey$ where axisymmetric flow dominates (see the analysis of turbulent puffs and localised invariant solutions in \citet{Ritter2018} at moderate $\Rey$). 

\section{Conclusion}
By using a technique combining a moving frame of reference and an artificial damping, we were able to simulate the fronts at high Reynolds numbers in short periodic pipes. The applicability of this technique confirms that the fronts are locally self-sustained at high Reynolds numbers. We measured the global propagation speed of turbulent fronts up to $\Rey=10^5$ in pipe flow, which is the highest Reynolds number considered so far to our best knowledge. Our results presented scalings of the front speeds with $\Rey$ in the widest $\Rey$ range so far, see (\ref{equ:scaling_DF}) and (\ref{equ:scaling_UF}). The scaling of the UF speed is very close to that proposed theoretically for a model system by \citet{Barkley2015}. The monotonically increasing trend of the DF speed is in stark contrast to the measurement of \citet{Wygnanski1973} above $\Rey\simeq 10000$, {\color{black} which was affected by the insufficient pipe length that didn't allow the flow to fully develop}.
Besides speed measurement, we also qualitatively discussed about the mechanism that determines the front speed, 
{\color{black}which can be summarized as the following points. 
\begin{enumerate}
\item A strong front can keep a characteristic shape and speed because there is transition to turbulence continually occurring at the tip of the fronts (see figure 7 and 8 and the supplementary movies). The speed of the fronts should be determined by the radial position of the transition at the front tips. 
\item Our energy budget analysis showed that the position of the transition at the front tip moves towards the wall at the upstream front and moves towards the pipe center at the downstream front, as figure 13 shows. This is consistent with the known fact that, as Re increases, the amplitude of perturbations needed to trigger pipe flow turbulence decreases.
\item The trend in the front position suggests the monotonic trend in the front speed. The closer to the pipe wall, the lower the local flow speed at the upstream front tip. Therefore, one can expect a lower convection speed of the transition-inducing disturbances so that a lower upstream front speed. On the contrary, the closer to the pipe center, the faster the local flow speed at the downstream front tip. Therefore, one can expect a faster convection of the transition-inducing disturbances so that a faster downstream front speed (approaching a limit).
\end{enumerate} }
Based on our analysis, we proposed that the speeds of both fronts would keep their respective monotonic trends as $\Rey\to\infty$. We also showed that the flow structures at the tips of the UF and DF, where local transition to turbulence continually occurs, are different. At the UF, the transition occurs in high shear region near the pipe wall and the dominant structures exhibit high azimuthal wave numbers, whereas the transition occurs in low shear region close to the pipe center at the DF, exhibiting low azimuthal wave numbers (but not axisymmetric). The different transition scenarios are possibly responsible for the asymmetry in the scaling of the two fronts. However, more quantitative studies are needed in the future for elucidating the instability and transition mechanisms at the front tip, which fundamentally determine the kinematics and flow structures of the fronts.

\section{Acknowledgements}
\indent K.C. and B.S. acknowledge financial support from the National Natural Science Foundation of China under grant number 91852105 and 91752113 and from Tianjin University under grant number 2018XRX-0027. \xu{D.X. acknowledges the partial support from NSFC Basic Science Center Program for ``Multiscale Problems in Nonlinear Mechanics'' (No. 11988102).} We acknowledge the computing resources from TianHe-2 at the National Supercomputer Centre in Guangzhou, where most of the simulations were performed. Part of the simulations were performed on TianHe-1(A) at the National Supercomputer Centre in Tianjin. We thank Bj\"orn Hof, Marc Avila and Dwight Barkley for insightful discussions on the topic over the years.

\section*{Conflict of interests}
The authors declare no conflict of interests.

\appendix
\section{Grid resolution}
\label{sec:appendix}

\begin{figure}
\centering
\includegraphics[width=0.9\linewidth]{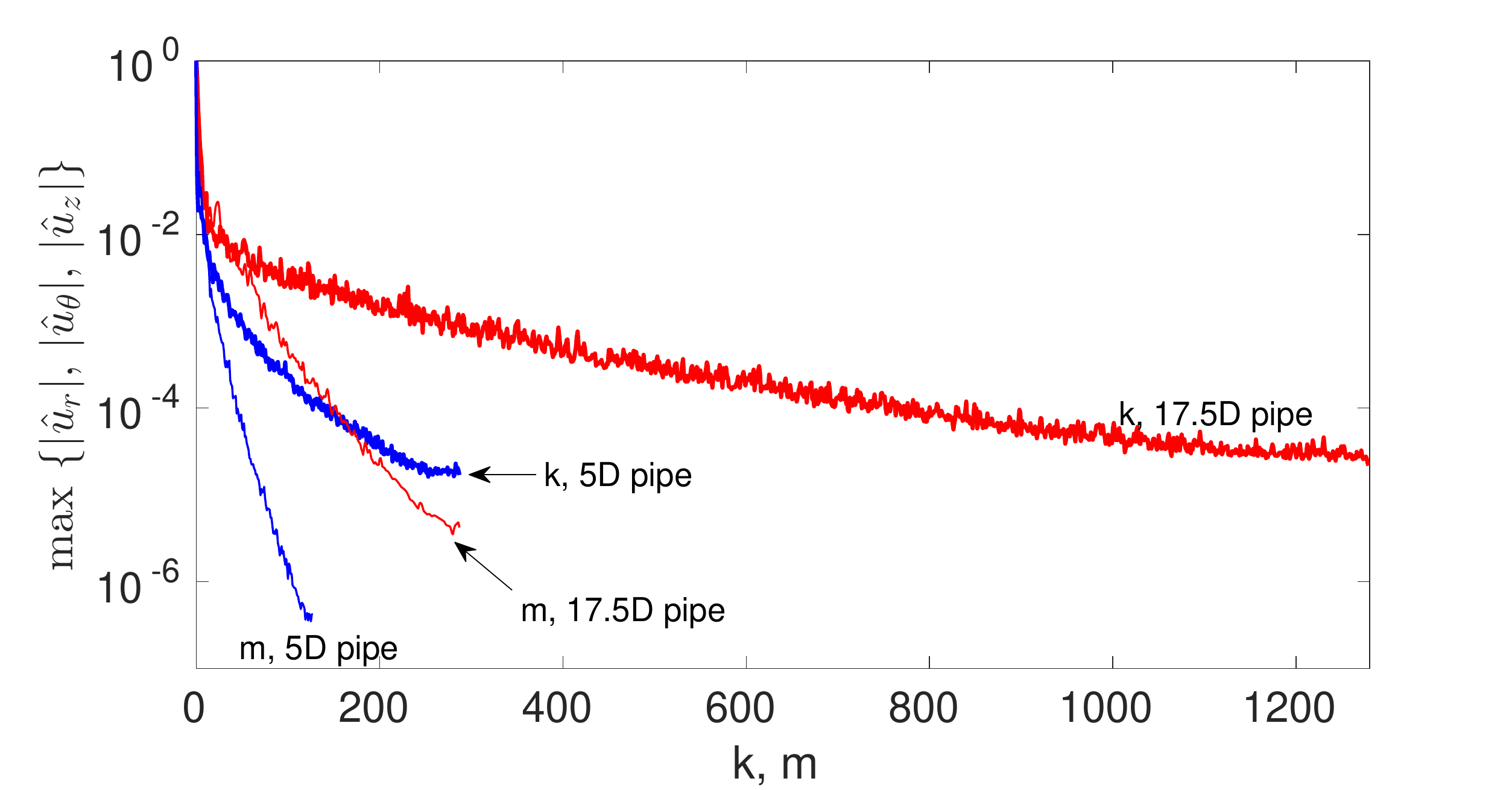}
\caption{\label{fig:spectra_Re25000} The velocity spectra for the DF in the $17.5D$ (red) and $5D$ (blue) pipes at $Re=25000$ using the resolutions shown in table \ref{tab:resolution_17.5D} and \ref{tab:resolution_5D}. The horizontal axis represents the index of the Fourier modes (streamwise mode $k$ or azimuthal mode $m$) and the vertical axis represents the modulus of the Fourier coefficient, maximised over three velocity components, radial direction and one of the two wavenumbers. Each data set is normalized by its maximum in order to compare the decrease in the spectra between different sets.}
\end{figure}

For the simulations in the $17.5D$ pipe, we used typical grid resolutions for DNS of fully turbulent flow (see e.g. \citet{Wu2008,Ahn2015}). Specifically, we used a grid spacing of $\Delta z^+\approx 7.8-9.6$ in the streamwise direction and $0.5D^+\Delta\theta\approx 5.6-8.6$ in the azimuthal direction at the pipe wall, in which the normalization is based on the viscous length unit evaluated for the fully turbulent flow at the respective Reynolds number. See table \ref{tab:resolution_17.5D} for the detail. These resolutions can assure a decrease of no less than four orders of magnitude from the lowest to the highest Fourier mode in the velocity spectra, see an example for $Re=25000$ in figure \ref{fig:spectra_Re25000}.

\begin{table}
\centering
\begin{tabular}{p{1cm}p{1.7cm}p{1.3cm}p{1.3cm}p{1.7cm}p{1.3cm}p{1.3cm}p{1.3cm}p{1.3cm}}
$\Rey$	& $\Rey_{\tau}=\frac{u_\tau D}{2\nu}$	& $\Delta r^+_{min}$	&  $\Delta r^+_{max} $	& $0.5D^+\Delta\theta$	& $\Delta z^+$	 & $N$ & $M$ & $K$\\
5000	& 344  	& 0.07	& 3.5	& 5.6	& 7.8 & 72 & 192 & 768\\
7500	& 490	& 0.06	& 3.8	& 8.0	& 8.4 & 96 & 192 & 1024\\
10000	& 630 	& 0.05	& 3.9	& 8.2	& 9.6 &  120 & 240 & 1152\\
17500	& 1026	& 0.03	& 4.0	& 8.4	& 9.3 & 192 & 384 & 1920\\
25000	& 1400	& 0.04	& 5.4	& 7.6 	& 9.6 & 216 & 576 & 2560 \\
40000	& 2114	& 0.04	& 6.6	& 8.6 	& 9.6 & 288 & 768 & 3840  \\
\end{tabular}
\caption{\label{tab:resolution_17.5D} Grid resolutions used for simulations in the $17.5D$ pipe. $\Rey_{\tau}$ and the viscous length unit $u_\tau/\nu$ are evaluated for the fully developed turbulent pipe flow at the same Reynolds numbers. {\color{black}Grid numbers in the radial, azimuthal and axial directions are given by $N$, $M$ and $K$, respectively.} Note that our $\Rey_{\tau}$ should be halved to compare with the literature in which usually the pipe radius is the reference length scale.}
\end{table}

\begin{table}
\centering
\begin{tabular}{p{1cm}p{1.7cm}p{1.1cm}p{1.1cm}p{1.6cm}p{1.4cm}p{1cm}p{1.5cm}p{1.5cm}}
$\Rey$	& $\Rey_{\tau}=\frac{u_\tau D}{2\nu}$	& $\Delta r^+_{min}$	&  $\Delta r^+_{max} $	& $0.5D^+\Delta\theta$	& $\Delta z^+$ & $N$ & $M$ & $K$	\\
17500	& 1026	& 0.03 	& 4.0	& 6.3 (16.8)	& 8.9 (8.9) & 192 & 512 (192) & 576 (576) \\
25000	& 1400	& 0.04	& 5.4 	& 6.9 (17.2)	& 9.1 (12.2) & 216 & 640 (256) & 768 (576)\\
40000	& 2114	& 0.04	& 6.6	& 6.9 (17.3)    & 9.2 (16.5) & 288 & 960 (384) & 1152 (640)\\
60000	& 3016	& 0.06	& 9.4	& 8.2 (24.7)    & 9.8 (15.7) & 384 & 1152 (384) & 1536 (960)\\
$10^5$	& 4714	& - (0.04) 	& -  (9.2)  	&  -  (28.9)   & - (20.5) &  - (384) &  - (512)  & -  (1024)
\end{tabular}
\caption{\label{tab:resolution_5D} Grid resolutions used for simulations in the $5D$ pipe. $\Rey_{\tau}$ and the viscous length unit are evaluated for the fully developed turbulent pipe flow at the same Reynolds numbers. {\color{black}Grid numbers in the radial, azimuthal and axial directions are given by $N$, $M$ and $K$, respectively.} The numbers in the parentheses are for the DF.}
\end{table}

The resolutions for simulations in the $5D$ pipe are listed in table \ref{tab:resolution_5D}. The resolutions for the UF are nearly the same as those used for the $17.5D$ pipe. However, the resolutions for the DF are drastically lowered, see the numbers in the parentheses. As we explained in the main text, since we only need to simulate the tip of the DF, which is located in the low shear region near the pipe center and exhibits larger flow structures than the small length scales in fully developed turbulent flow, we can resolve the flow using much lower resolution in both streamwise and azimuthal directions. 
Figure \ref{fig:spectra_Re25000} shows the comparison between the velocity spectra of a flow field for the DF simulated in the $17.5D$ and $5D$ pipes at $Re=25000$. 
As can be seen, both resolutions give a decrease of more than four orders of magnitude in the velocity spectra and even a lower resolution could have been used for the simulation in the $5D$ pipe to obtain an equally well resolved flow field as that in the $17.5D$ pipe.

\bibliographystyle{jfm}

\end{document}